\documentclass[aps,prd,nofootinbib,preprint,eqsecnum,superscriptaddress,12pt]{revtex4}

\usepackage{epsfig}
\usepackage{graphicx}
\usepackage{dcolumn}
\usepackage{amsmath}
\usepackage{enumerate}

\def\beq{\begin{equation}}
\def\eeq{\end{equation}}

\newcommand{\capdef}{}
\newcommand{\mycaption}[2][\capdef]{\renewcommand{\capdef}{#2}%
       \caption[#1]{{\footnotesize #2}}}
\makeatletter
\renewcommand{\fnum@table}{\textbf{\tablename~\thetable}}
\renewcommand{\fnum@figure}{\textbf{\figurename~\thefigure}}
\makeatother
\def\ltap{\ \raisebox{-.4ex}{\rlap{$\sim$}} \raisebox{.4ex}{$<$}\ }
\def\gtap{\ \raisebox{-.4ex}{\rlap{$\sim$}} \raisebox{.4ex}{$>$}\ }

\newcounter{myenumi}

\renewcommand{\themyenumi}{\roman{myenumi}}
{\end{list}}

\newlength{\myem}
\newcounter{mysubequation}[equation]

\def\ltap{\ \raisebox{-.4ex}{\rlap{$\sim$}} \raisebox{.4ex}{$<$}\ }
\def\gtap{\ \raisebox{-.4ex}{\rlap{$\sim$}} \raisebox{.4ex}{$>$}\ }

\begin{document}

\renewcommand{\thefootnote}{\alph{footnote}}

\begin{flushright} 
CERN-PH-TH/2005-105\\
SISSA 34/2005/EP\\
\end{flushright}

\vspace*{1cm}

\renewcommand{\thefootnote}{\fnsymbol{footnote}}
\setcounter{footnote}{-1}

\title{
\vskip12pt~\\
Explaining LSND by a decaying sterile neutrino\\[5mm]}

\author{Sergio Palomares-Ruiz}
\email{sergio.palomares-ruiz@vanderbilt.edu} 
\affiliation{Department of Physics and Astronomy, 
             Vanderbilt University, Nashville, TN 37235, USA}

\author{Silvia Pascoli} \email{Silvia.Pascoli@cern.ch}
\affiliation{Physics Department, Theory Division, CERN, 
             CH--1211 Geneva 23, Switzerland}

\author{Thomas Schwetz} \email{schwetz@sissa.it}
\affiliation{Scuola Internazionale Superiore di Studi Avanzati,
             Via Beirut 2--4, I--34014 Trieste, Italy
\vspace*{1cm}
}

\begin{abstract}
\vspace*{0.5cm} 
  We propose an explanation of the LSND evidence for electron
  antineutrino appearance based on neutrino decay. We introduce a
  heavy neutrino, which is produced in pion and muon decays because of
  a small mixing with muon neutrinos, and then decays into a scalar
  particle and a light neutrino, predominantly of the electron
  type. We require values of $g m_4\sim$~few~eV, $g$ being the
  neutrino--scalar coupling and $m_4$ the heavy neutrino mass,
  e.g.\ $m_4$ in the range from 1~keV to 1~MeV and $g \sim
  10^{-6}$--$10^{-3}$. Performing a fit to the LSND data as well as
  all relevant null-result experiments, we show that all data can be
  explained within this decay scenario. In the minimal version of the
  decay model, we predict a signal in the upcoming MiniBooNE
  experiment corresponding to a transition probability of the same
  order as seen in LSND. In addition, we show that extending our model
  to two nearly degenerate heavy neutrinos it is possible to introduce
  CP violation in the decay, which can lead to a suppression of the
  signal in MiniBooNE running in the neutrino mode. We briefly discuss
  signals in future neutrino oscillation experiments, we show that our
  scenario is compatible with bounds from laboratory experiments, and
  we comment on implications in astrophysics and cosmology.
\end{abstract}

\maketitle

\renewcommand{\thefootnote}{\arabic{footnote}}
\setcounter{footnote}{0}

\section{Introduction}

The LSND experiment at LANSCE in Los Alamos took data from
1993--1998 and observed an excess of $\bar\nu_e$~\cite{Aguilar:2001ty}.
This was interpreted as evidence of $\bar\nu_\mu \rightarrow
\bar\nu_e$ transitions. The KARMEN experiment at the spallation source
ISIS in England was looking at the same appearance chanel in the
years 1997--2001 at a slightly different baseline than LSND, but no
positive signal was found~\cite{karmen}.
Reconciling the evidence for $\bar\nu_\mu\to\bar\nu_e$ appearance
observed in LSND with the other evidence for neutrino
oscillations~\cite{sk-atm,dip,Aliu:2004sq,sno,Araki:2004mb} remains a
challenge for neutrino phenomenology.  It turns out that introducing a
fourth sterile neutrino~\cite{sterile} does not lead to a satisfactory
description of all data in terms of neutrino
oscillations~\cite{Maltoni:2002xd,strumia} (see
Ref.~\cite{Maltoni:2004ei} for a recent update) because of tight
constraints from atmospheric~\cite{sk-atm}, solar~\cite{sno}, and
null-result short-baseline (SBL)
experiments~\cite{karmen,Dydak:1983zq,Declais:1994su,Astier:2003gs}.
In view of this, several alternative explanations have been proposed,
some of them involving very speculative physics: four-neutrino
oscillations plus neutrino decay~\cite{Ma:1999im}, three neutrinos and
CPT violation~\cite{cpt,strumia}, a lepton number violating muon
decay~\cite{LNV}, five-neutrino oscillations~\cite{Sorel:2003hf}, four
neutrinos and CPT violation~\cite{4nu-cpt}, a
low-reheating-temperature Universe~\cite{GPRP}, CPT-violating quantum
decoherence~\cite{Barenboim:2004wu}, mass-varying
neutrinos~\cite{MVN}, and shortcuts of sterile neutrinos in extra
dimensions~\cite{Pas:2005rb}.  A critical test of the LSND signal will
come soon from the MiniBooNE experiment~\cite{miniboone}, which is
looking for $\nu_\mu\to\nu_e$ appearance in a similar range of
$L/E_\nu$ as LSND, $L$ being the distance travelled by the neutrinos
and $E_\nu$ their energy.

In this work we assume the existence of heavy (mainly sterile)
neutrinos, $n_h$, with a small mixing with the muon neutrino. We
denote their masses by $m_h$. The LSND signal is explained through the
decay of $n_h$ into a light neutrino, mixed predominantly with the
electron neutrino. A natural way to introduce neutrino decays is by
means of a term in the Lagrangian, which couples neutrinos to a light
scalar, similar to the so-called Majoron
models~\cite{majoronmodels}. This term can be related to the neutrino
mass generation mechanism through the spontaneous breaking of a lepton
number symmetry via a non-zero vacuum expectation value of the scalar
field. Similarly to these models, we introduce an interaction term
between $n_h$, light neutrinos and a scalar.  However, as we are only
interested in the phenomenological consequences of such a term for the
LSND experiment as well as for the other relevant neutrino
experiments, we assume that this term arises at low energy as an
effective interaction, not necessarily related to neutrino mass
generation. We show that in the simplest case of one heavy sterile
neutrino, $n_4$, for $g m_4 \sim 1$--$10$ eV, e.g. neutrino masses
around 100~keV and neutrino--scalar couplings $g \sim 10^{-5}$, LSND
data can be explained in complete agreement with the null-results of
other SBL neutrino experiments.

Light neutrino decay has been considered as an alternative solution to
neutrino oscillations, to explain the atmospheric neutrino
anomaly~\cite{nudecay,atmosdecay} and the solar neutrino
puzzle~\cite{nudecay,solardecay}. Although the recent observation of an
oscillatory behaviour in atmospheric neutrino~\cite{dip} and
KamLAND~\cite{Araki:2004mb} data excludes a pure decay scenario as the
possible explanation of these problems, a combined scenario of
neutrino oscillations and decay is not excluded yet.  
We do not consider these possibilities here and we
assume that atmospheric, solar and KamLAND neutrino
transitions are explained by neutrino oscillations.

The decay of an exotic neutral massive particle, called karmino,
produced in pion decays was considered as a possible explanation for
an anomaly seen in the first data set of the KARMEN
experiment~\cite{karmen1}, which however, was not subsequently
confirmed~\cite{karmen2}. This hypothetical particle would move
non-relativistically ($\beta \sim 0.02$) from the source to the
detector.  If it were a massive neutrino, strong bounds on the mixing
and lifetime could be placed from anomalous pion and muon
decays~\cite{boundsdecay} (for a recent review of the bounds, see
Ref.~\cite{KPS04}).  Differently from this case, where no specific
decay scenario was adopted, we assume here a Majoron-like type of
decay. Furthermore, for typical values of masses, the heavy neutrino
is relativistic at the energies in the LSND and KARMEN experiments,
and we require it to decay before reaching the detector, with a
lifetime much smaller than the one of the karmino.
 
An explanation of the LSND result invoking the interplay between
neutrino oscillations and neutrino decay has been considered
previously in Ref.~\cite{Ma:1999im}. A fourth
neutrino is introduced with a mass of a few eV, which decays into a
massless Goldstone boson (the Majoron) and into light neutrinos. The
signal in LSND is provided by mixing, similar to (3+1) four-neutrino
oscillations, whereas the decay serves to circumvent the constraints
from the CDHS experiment. This approach requires very large couplings
$g \sim \mathcal{O}(10^3)$ and a mixing of $\nu_e$ with the heavy mass
state, and hence it appears to be in conflict with various bounds on
neutrino--scalar couplings (see Section~\ref{sec:cosmo} for a
discussion).  In contrast, in our model, the signal in the LSND
experiment is provided entirely by the decay, and no mixing of $\nu_e$
with the heavy neutrino is required. This allows us to invoke neutrino
masses in the 100~keV range, which in turn permits a small coupling of
$g\sim 10^{-5}$, in agreement with existing bounds.

The outline of the paper is as follows. In Section~\ref{sec:framework}
we present the minimal decay framework and deduce the probabilities,
which are needed to calculate the event rates. In
Section~\ref{sec:LSND} we present our analysis of LSND and KARMEN data 
within this scenario, and in Section~\ref{sec:SBL} we show that the
decay framework is not in contradiction with any other null-result
experiment by performing a combined analysis; we compare the
quality of the fit to the one for oscillations. In
Section~\ref{sec:mini} we discuss the prediction for MiniBooNE and
other future neutrino experiments within the minimal decay framework,
whereas in Section~\ref{sec:CP} we show that, allowing for, at least,
two sterile neutrinos and complex couplings, it is possible to
obtain a CP-violating interference between oscillation and decay
amplitudes, which would suppress the neutrino signal in MiniBooNE, but
at the same time provide the antineutrino signal in LSND. In
Section~\ref{sec:cosmo}, we argue that our model is consistent with
existing bounds from laboratory experiments, and we comment on
implications for cosmology and astrophysics. In
Section~\ref{sec:conclusions} we present our final remarks.

\section{Decay framework and transition probabilities}
\label{sec:framework}

Let us consider the general case of $N$ Majorana neutrinos. We take
the three light neutrinos responsible for solar and atmospheric
oscillations to be $\nu_{1,2,3}$ with masses $m_{1,2,3} \lesssim
1$~eV, while the heavy neutrinos have masses $m_{4,5,\ldots}\gg
m_{1,2,3}$.  Furthermore we assume for the scalar mass $m_{1,2,3}
\lesssim m_\phi \ll m_{4,5,\ldots}$, such that the three light
neutrinos are stable. Hence the terms in the Lagrangian relevant to
the decay (with scalar couplings\footnote{The case of pseudo-scalar
couplings gives exactly the same results in the relativistic
approximation we will use. It can also be shown that for processes in
which all of the involved stares are on-shell, a derivative
interaction term can be rewritten in pseudo-scalar
form~\cite{Farzan02}.}) are given, in the mass basis, by
\beq\label{eq:Ldec}
\mathcal{L} = - 
\sum_{l,h} g_{hl} \, \overline{\nu}_{lL} \, n_{hR} \, \phi  +
\mathrm{h.c.}\,, 
\eeq
where here and in the following the ranges for the indexes are 
$l = 1,2,3$ and $h = 4,5,\ldots$. In general the
coupling matrix $g_{hl}$ will be complex. 

In the case of Majorana particles, neutrinos are identical to
antineutrinos. Weak interactions couple to left-handed (chiral)
neutrinos and right-handed (chiral) antineutrinos while the
propagation states are those of definite helicity. In the relativistic
case, where helicity and chirality are approximately the same, one can
identify neutrinos and antineutrinos with helicity states up to terms
of order $m/E_\nu$.  It follows from Eq.~(\ref{eq:Ldec}) that not only
the decay $n \to \bar{\nu} + \phi$ can occur, but also $n \to \nu +
\phi$ is possible~\cite{Kim:1990km}. Both processes are suppressed by
terms of order $m_h/E_{n_h}$.  For the former process the suppression
is due to angular momentum conservation, while for the latter the
chirality-flipping nature of the interaction, Eq.~(\ref{eq:Ldec}), is
responsible for the suppression.\footnote{Throughout this work we will
assume Majorana neutrinos. Here we just remark that in the case of
Dirac neutrinos the decay is analogous. However, the decay products of
the helicity-flipping channel are unobservable, since they are the
right-(left-)handed component of the neutrino (antineutrino) fields,
which do not participate in weak interactions.}

In the approximation $m_l \approx 0$, $m_\phi \approx 0$ and $m_h \ll
E_{n_h}$, the differential decay rate of an $n_h$ of helicity $r$
with energy $E_{n_h}$ into a $\nu_l$ of helicity $s$ with energy
$E_{\nu_l}$ is
\beq\label{dkrate}
\frac{d\Gamma_{n_h^r\to \nu_l^s}(E_{n_h})}{dE_{\nu_l}} = 
\frac{1}{16 \pi E_{n_h}^2} \, |\mathcal{M}_{rs}|^2  \,,
\eeq
where the indices $r,s$ take the values `$-$' for neutrinos and `$+$'
for antineutrinos, and the matrix element is given by\footnote{
According to angular momentum conservation the matrix element
Eq.~(\ref{matrixel}) can be written also as $|\mathcal{M}_{rs}|^2 =
|g_{hl}|^2 \, m_h^2 (1\mp\cos\theta)/2$ for $r=s$ ($r\neq s$), where
$\theta$ is the angle between the spin of $n_h$ and the direction of
$\nu_l$ in the rest frame of $n_h$.}
\beq\label{matrixel}
|\mathcal{M}_{rs}|^2 = |g_{hl}|^2 \, m_h^2 \times
\left\{
\begin{array}{l@{\quad}l}
E_{\nu_l}/E_{n_h} & r=s \\
(1 - E_{\nu_l}/E_{n_h}) & r \neq s 
\end{array} \right. \,.
\eeq
For relativistic neutrinos, the limiting
values of the final neutrino energy are $0 < E_{\nu_l} < E_{n_h}$, and
hence, the partial decay rate for $n_h^r \to \nu_l^s + \phi$ for {\it
both} cases, helicity-flipping ($r\neq s$) and helicity-conserving ($r
 = s$), is the same and given by~\cite{Kim:1990km}
\beq\label{dk}
\Gamma_{hl} = \Gamma_{n_h^r \to \nu_l^s} =
\frac{|g_{hl}|^2 \, m_h^2}{32 \pi E_{n_h}} \, .
\eeq
The total decay rate of $n_h$ is then given by 
$\Gamma_h = 2 \sum_l \Gamma_{hl}$.

The flavour neutrinos $\nu_\alpha$ ($\alpha = e, \mu, \tau, s_1,
s_2,\ldots$) are related to the massive neutrinos in
Eq.~(\ref{eq:Ldec}) by 
\beq
\nu_\alpha = \sum_l U_{\alpha l} \nu_l + \sum_h U_{\alpha h} n_h \,.
\eeq
The differential
probability that an (anti)neutrino of flavour $\alpha$ with energy
$E_{\nu_\alpha}$ is converted into an (anti)neutrino of flavour
$\beta$ with energy in the interval $[E_{\nu_\beta}, E_{\nu_\beta} +
dE_{\nu_\beta}]$ is:   
\begin{eqnarray}
\frac{dP_{\nu_\alpha^r\to\nu_\beta^s}(E_{\nu_\alpha})}{dE_{\nu_\beta}}
&=& 
  \left| \sum_l U_{\beta l}^{(s)} U_{\alpha l}^{(r)*}
  \exp\left(-i\frac{m_l^2 L}{2E_{\nu_\alpha}} \right) \right|^2 
  \delta(E_{\nu_\alpha} - E_{\nu_\beta}) \, \delta_{rs}\nonumber\\
&+&
  \left| \sum_h U_{\beta h}^{(s)} U_{\alpha h}^{(r)*} 
  \exp\left(-i\frac{m_h^2 L}{2E_{\nu_\alpha}} \right) 
  \exp\left(-\frac{\Gamma_h L}{2} \right) \right|^2 
  \delta(E_{\nu_\alpha} - E_{\nu_\beta}) \, \delta_{rs}\nonumber\\
&+&
  W_{rs}(E_{\nu_\alpha}, E_{\nu_\beta}) \int_0^L dL'
  \frac{dP_{\nu_\alpha^r\to \nu_\beta^s}^\mathrm{dec}}{dL'} \,,
  \label{eq:P1}
\end{eqnarray}
where we have introduced the notation $X^{(r)} \equiv X$ for $r=-$ and
$X^{(r)} \equiv X^*$ for $r=+$. In Eq.~(\ref{eq:P1}), we have used
the fact that the light neutrinos $\nu_l$ are not coherent with $n_h$ 
because of the large mass difference,
while both $\nu_l$ and $n_h$ are assumed to be coherent
among themselves. The first term describes the oscillations of light
neutrinos, whereas the second one describes oscillations of the heavy
neutrinos weighted by the probability that they do not decay. 
These oscillation terms
are only present in the helicity-conserving channel ($r=s$). The third term
describes the appearance of decay products, and is present for the
helicity-conserving as well as helicity-flipping channels. Here
$W_{rs}(E_{\nu_\alpha}, E_{\nu_\beta})$ is the normalised energy
distribution of the decay products:
\beq\label{eq:W}
W_{rs}(E_{\nu_\alpha},E_{\nu_\beta}) \equiv
\frac{1}{\Gamma_{hl}}
\frac{d\Gamma_{n_h^r\to \nu_l^s}(E_{\nu_\alpha})}{dE_{\nu_\beta}}
= 
2 \, \Theta(E_{\nu_\alpha} - E_{\nu_\beta}) \times
\left\{
\begin{array}{l@{\quad}l}
E_{\nu_\beta}/E_{\nu_\alpha}^2 & r=s , \\
(E_{\nu_\alpha} - E_{\nu_\beta})/E_{\nu_\alpha}^2 & r \neq s ,
\end{array} \right.
\eeq
where $\Gamma_{hl}$ is the partial decay rate for $n_h\to\nu_l$ given
in Eq.~(\ref{dk}). The relativistic approximation $m_l \approx 0$
allows us to factor out $W_{rs}(E_{\nu_\alpha}, E_{\nu_\beta})$ in
Eq.~(\ref{eq:P1}). The term $dP_{\nu_\alpha^r\to\nu_\beta^s}^\mathrm{dec} /
dL$ is the differential probability that the heavy component of the
neutrino $\nu^r_\alpha$ decays at the distance $[L,L+dL]$ and the
decay product interacts as a $\nu^s_\beta$. Adopting the effective
operator method of Ref.~\cite{LOW01} it can be calculated as
\beq\label{eq:P_diff}
\frac{dP_{\nu_\alpha^r\to\nu_\beta^s}^\mathrm{dec}}{dL} =
\left| \sum_l U_{\beta l}^{(s)} \sum_h U_{\alpha h}^{(r)*} \,
\mathcal{A}_{hl}^{(r)} \, \exp\left(-i\frac{m_h^2 L}{2E_{\nu_\alpha}}
\right) \exp\left(-\frac{\Gamma_h L}{2} \right) \right|^2 .
\eeq
Here $\mathcal{A}_{hl}$ is an effective amplitude for the decay of
$n_h$ into $\nu_l$, similar to the ``appearance operator'' of
Ref.~\cite{LOW01}, and it is given by
\beq\label{eq:app_amplitude}
\mathcal{A}_{hl} = g_{hl} A_h \,,\qquad 
A_h \equiv \frac{\sqrt{\Gamma_{hl}}}{|g_{hl}|}\,.
\eeq
Note that in the limit where the light neutrino masses can be
neglected in the decay kinematics, the real quantity $A_h$ is
independent of the index $l$. Since for SBL experiments we have
$\Delta m^2_{ll'} L / E \ll 1$ we can neglect oscillations of the
decay products to derive Eq.~(\ref{eq:P_diff}), but we include the
possibility of oscillations of the heavy states before they decay.

In the simplest case in which $N=4$, Eq.~(\ref{eq:P_diff}) becomes
\beq\label{eq:P_diff2}
\frac{dP_{\nu_\alpha^r\to\nu_\beta^s}^\mathrm{dec}}{dL} =
|U_{\alpha 4}|^2 |g_\beta^{rs}|^2 \, A_4^2 \, e^{-\Gamma_4 L}  \,,
\eeq
where we have defined the couplings in the flavour basis by 
\beq\label{eq:g_flavour}
g_\beta^{rs}
\equiv \sum_l U_{\beta l}^{(s)} g_{4l}^{(r)}\,. 
\eeq

\section{LSND and KARMEN}
\label{sec:LSND}

Let us now calculate the $\nu_\mu^r \to \nu_e^s$ appearance
probability relevant to the LSND and KARMEN experiments. Unlike other
explanations of the LSND result based on sterile neutrinos, our model
does not require a mixing of the electron neutrino with the heavy mass
state. Therefore, for simplicity we assume $U_{e4} = 0$. This implies
that the first two lines in Eq.~(\ref{eq:P1}) disappear and using
Eq.~(\ref{eq:P_diff2}) we find
\beq\label{eq:P2}
\frac{dP_{\nu_\mu^r\to \nu_e^s}(E_{\nu_\mu})}{dE_{\nu_e}} =
W_{rs}(E_{\nu_\mu}, E_{\nu_e}) \, \frac{1}{2} \,
|U_{\mu 4}|^2 \, R_e
\, (1 - e^{-\Gamma_4 L}) \,,
\eeq
where we have defined the branching ratios by
\begin{equation}
R_{\alpha} \equiv \frac{|g_\alpha|^2}{\bar g^2} \,,\quad
\bar g^2 \equiv \sum_l |g_{4l}|^2 = 
\sum_{\alpha = e,\mu,\tau,s} |g_\alpha|^2 \,.
\end{equation}
We have used $\Gamma_4 = 2 \bar g^2 A^2_4$, and for simplicity we
neglect the dependence of $g_\alpha$ on the helicity indices, i.e. we
neglect complex phases of $U_{\alpha l}$ and $g_{4l}$ (compare to 
Eq.~(\ref{eq:g_flavour})). The factor $1/2$ in Eq.~(\ref{eq:P2})
accounts for the fact that half of the initial neutrinos decay into
neutrinos and half of them into antineutrinos.

In LSND, neutrinos are produced by the decay at rest (DAR) of pions
$\pi^+ \to \mu^+ + \nu_\mu$ and by the subsequent decay of the muons
$\mu^+ \to e^+ + \nu_e + \bar\nu_\mu$. 
In the detector, 
the appearance of $\bar{\nu}_e$ 
is searched for by using the reaction $\bar\nu_e + p
\to e^+ + n$. 
In case of oscillations, only the $\bar\nu_\mu$ from the
muon decay contribute to the signal.
In the decay scenario, in
addition, the $\nu_\mu$ from the primary pion decay will give a
contribution to the $\bar\nu_e$ signal as well. 
Note that the $\nu_e$ from the
muon decay will not contribute to the signal because of our assumption
$U_{e4}=0$. The pion decay leads to a mono-energetic $\nu_\mu$ beam with
an energy of $E_{\nu_\mu}^{(\pi)} = 29.8$~MeV, whereas the muon decay
gives $\bar\nu_\mu$ with a continuous spectrum,
$\phi_\mu(E_{\bar\nu_\mu})$, rising up to
$E_{\bar\nu_\mu}^\mathrm{max} = 52.8$~MeV.
The total number of
neutrinos is in both cases $\phi_0 = \int dE_{\bar\nu_\mu}
\phi_\mu(E_{\bar\nu_\mu})$. The expected number of $\bar\nu_e$ events
with neutrino energy in the interval $[E_{\bar\nu_e} , E_{\bar\nu_e}+ 
dE_{\bar\nu_e}]$ is given by 
\beq\label{eq:N}
\frac{dN}{dE_{\bar\nu_e}} = C \, \sigma(E_{\bar\nu_e}) \, 
\left[
\phi_0 
\frac{dP_{\nu_\mu \to \bar\nu_e}(E_{\nu_\mu}^{(\pi)})}{dE_{\bar\nu_e}}
+
\int_{E_{\bar\nu_e}}^{E_{\bar\nu_\mu}^\mathrm{max}}
dE_{\bar\nu_\mu} \, \phi_\mu(E_{\bar\nu_\mu}) \,
\frac{dP_{\bar\nu_\mu \to \bar\nu_e}(E_{\bar\nu_\mu})}{dE_{\bar\nu_e}}
\right]\,,
\eeq
where $\sigma(E_{\bar\nu_e})$ is the detection cross section and $C$
is an overall constant containing the number of target particles,
efficiencies and geometrical factors. To calculate the observed number
of events, we fold Eq.~(\ref{eq:N}) with the energy resolution of the
detector and integrate over the relevant energy interval.

For the statistical analysis of LSND data we use the total number of
signal events implied by the transition probability $P=(0.264 \pm
0.040)\%$~\cite{Aguilar:2001ty}.
We have chosen an error such
that we can reproduce Fig.~6 of Ref.~\cite{Church:2002tc}, which is
based on an event-by-event analysis of the DAR data. Furthermore we
include spectral data in the form of 11 data points given in Fig.~24 of
Ref.~\cite{Aguilar:2001ty}, where the beam excess is shown as a
function of $L/E_\nu$. We fit these data with a free normalisation in
order to avoid a double counting of the information given already by
the total number of events. In our analysis we do not use the data
from decay in flight (DIF) neutrinos, since the appearance signal in
these data is less significant than in the DAR sample. Moreover, since
the total number of DIF neutrinos is at most two orders of magnitude
less than the $\bar\nu_\mu$ flux from DAR, one does not expect a
relevant contribution to the $\bar\nu_e$ appearance signal from DIF
due to the helicity-flipping decay.

\begin{figure}[t]
\centering 
\includegraphics[width=0.56\textwidth]{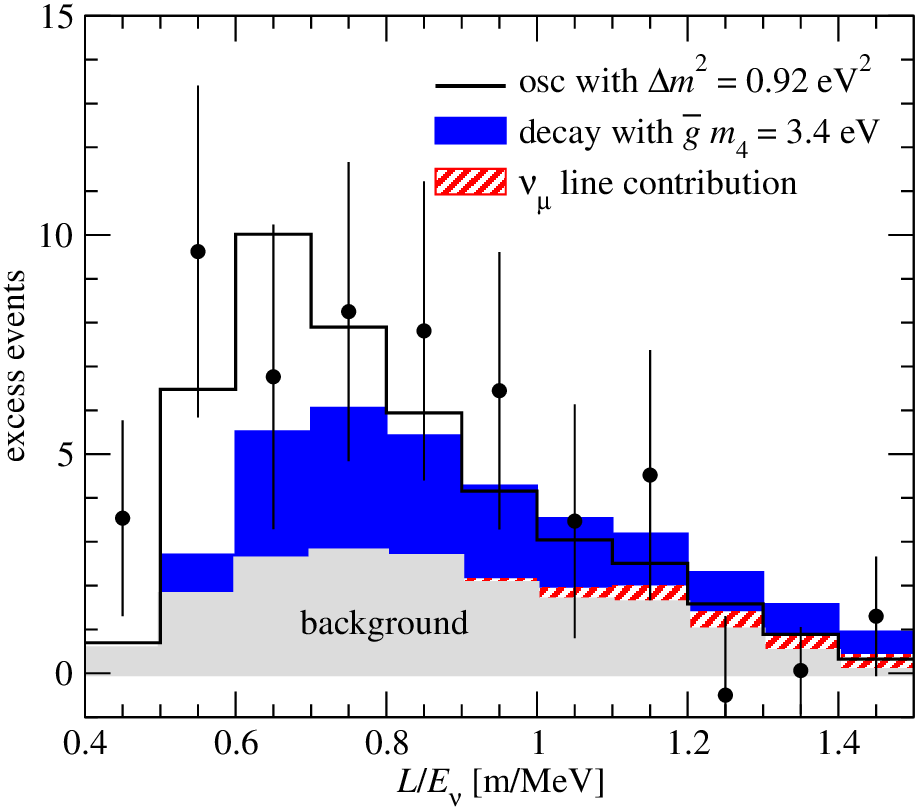}
  \mycaption{$L/E_\nu$ spectrum for LSND for decay and oscillations
  compared with the data given in Fig.~24 of
  Ref.~\cite{Aguilar:2001ty}. The hatched histogram shows the 
  contribution of the $\nu_\mu$ line from the primary pion decay in
  the decay scenario.}
\label{fig:spectrum}
\end{figure}

Our analysis of LSND data gives a best fit value of
$\chi^2_\mathrm{min} = 5.6/9$~dof for oscillations and
$\chi^2_\mathrm{min} = 10.8/9$~dof for decay. The reason for the
slightly worse fit for decay is the spectral distortion implied by the
energy distribution of the decay products given in Eq.~(\ref{eq:W}). In
Fig.~\ref{fig:spectrum} we compare the best fit spectra for
oscillations and decay with the data. In the decay
scenario we predict slightly too many events at low energies and too
few events in the high energy part of the spectrum. Note that the
contribution of the $\nu_\mu$ line from the primary pion decay is
rather small. This follows from the fact that in the helicity-flipping
decay mode, most of the decay products have a very small energy
(see Eq.~(\ref{eq:W})), which implies that they fall below the
detection threshold.\footnote{Note that here we neglected the
dependence of the couplings $g_\alpha$ on the helicity indices due to 
complex phases in $U_{\alpha l}$ and $g_{4l}$ according to
Eq.~(\ref{eq:g_flavour}). If this additional freedom is taken into 
account, different branching ratios $R_e$ for the helicity-conserving
and helicity-flipping decays can be obtained, which will modify the 
relative size of the $\nu_\mu$ line contribution.} Although the fit
in the decay scenario is slightly worse than for oscillations, the
overall goodness of fit (GOF) is still acceptable. The $\chi^2$ value
cited above implies a GOF of 29\%. Note, however, that a more refined
analysis of the LSND spectral data might allow a significant
discrimination between oscillations and decay. In the following we
will assume that LSND data can be explained by our decay model 
and we shall proceed with the combination with other SBL data. 

\begin{figure}[t]
\centering 
\includegraphics[width=0.95\textwidth]{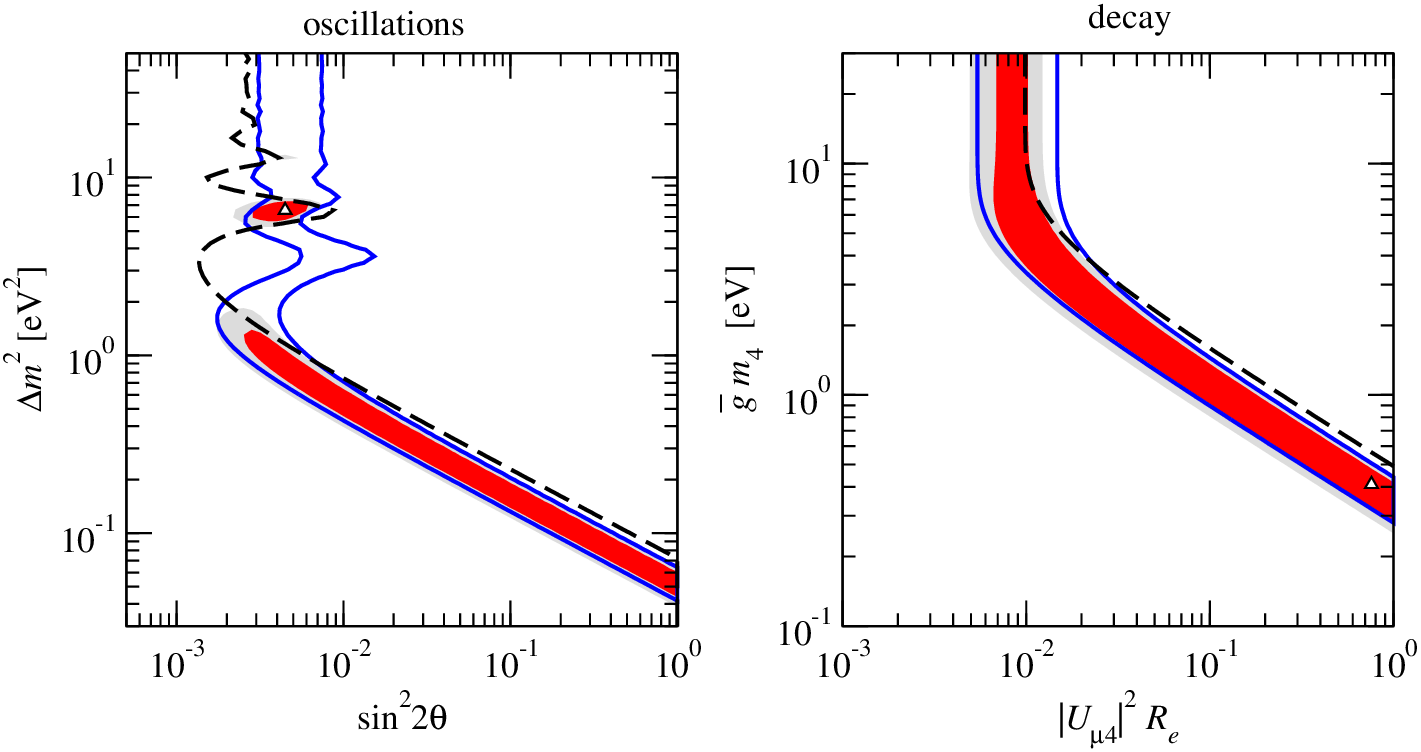}
  \mycaption{Allowed regions for LSND (solid) and KARMEN (dashed) at
  99\%~CL, and LSND and KARMEN combined (shaded regions) at 90\% and
  99\%~CL. The left panel corresponds to neutrino oscillations and the
  right panel to the decay scenario presented in this work.} 
\label{fig:lsnd-vs-karmen}
\end{figure}

First we discuss the compatibility of LSND and KARMEN in the decay
scenario (see Ref.~\cite{Church:2002tc} for a detailed analysis in the
case of oscillations). Much as in LSND, also in KARMEN neutrinos are
produced by the decay of pions at rest and the subsequent muon decay.
However, in KARMEN detailed time information is available from the
pulse structure of the proton beam and the excellent time resolution of
the detector. Taking into account the muon lifetime of $2.2\,\mu$s
compared with the pion lifetime of 26~ns, a possible contribution of
neutrinos from the pion decay was suppressed in the
$\bar\nu_\mu\to\bar\nu_e$ oscillation analysis by an appropriate 
time cut. Hence, in contrast to LSND the helicity-flipping decay of
neutrinos from the $\nu_\mu$ line does not contribute to the
$\bar\nu_e$ signal in KARMEN, and the first term in Eq.~(\ref{eq:N})
is absent. For the statistical analysis of KARMEN data we are using
the 9 data points as well as the expected background for the prompt
energy given in Fig.~11b of Ref.~\cite{karmen}.

In Fig.~\ref{fig:lsnd-vs-karmen} we show the allowed regions for LSND,
KARMEN, and the combination of both experiments for the case of
oscillations and for the decay scenario.  The comparison of the left
and right panels of Fig.~\ref{fig:lsnd-vs-karmen} shows that the
compatibility of LSND and KARMEN in the decay scenario is at the same
level as for oscillations.  The reason is that both the oscillation
phase, and the exponent $\Gamma_4L$ in Eq.~(\ref{eq:P2}), have the same
dependence on $L/E_\nu$.  This allows us to accommodate both the
positive signal in the LSND experiment and the KARMEN null result, by
taking into account the different baselines ($L_\mathrm{LSND} = 35$~m,
$L_\mathrm{KARMEN} = 18$~m). In addition, as discussed above, the
$\nu_\mu$ line gives a small contribution to the LSND signal, which is
absent in KARMEN because of the time cut.

A powerful tool to evaluate the compatibility of different data sets
is the parameter goodness of fit (PG) criterion discussed in
Ref.~\cite{Maltoni:2003cu}. It is based on the $\chi^2$ function
\beq\label{eq:PG}
\chi^2_\mathrm{PG} = 
\chi^2_\mathrm{tot,min} - \sum_i \chi^2_{i,\mathrm{min}} \,,
\eeq
where $\chi^2_\mathrm{tot,min}$ is the $\chi^2$ minimum of all data
sets combined and $\chi^2_{i,\mathrm{min}}$ is the minimum of the data
set $i$. Applying this method to the case of LSND and KARMEN we find
$\chi^2_\mathrm{PG} = 5.02$ for oscillations and $\chi^2_\mathrm{PG} =
4.97$ for decay. These $\chi^2$ numbers have to be evaluated for 2~dof,
corresponding to the two parameters in common to the two data sets
($\sin^22\theta$ and $\Delta m^2$ for oscillations and $|U_{\mu 4}|^2
R_e$ and $\bar g m_4$ for decay). This yields a PG of 8.1\% for
oscillations and 8.3\% for decay.

\section{Combined analysis of LSND and null-result experiments}
\label{sec:SBL}

Let us now discuss the survival probabilities relevant to the
analysis of SBL disappearance experiments. Under the assumption
$U_{e4}=0$ in Eq.~(\ref{eq:P1}),  no $\bar\nu_e$
disappearance is expected in SBL reactor experiments: $P_{\bar\nu_e\to
\bar\nu_e} = 1$. However, since we need mixing of $\nu_\mu$ with the
heavy states in order to explain the LSND signal, one expects some
effect in $\nu_\mu$ disappearance experiments. To simplify the
analysis, we will neglect in the following the appearance of $\nu_\mu$
in the decay products, i.e. we adopt the choice $R_e \approx 1$ and
$R_{\mu,\tau} \approx 0$ for the branching ratios of the decay. Then
Eq.~(\ref{eq:P1}) yields for the SBL $\nu_\mu$ survival probability 
\beq\label{eq:mu-surv}
P_{\nu_\mu \to \nu_\mu}^\mathrm{SBL,dec} = 
(1 - |U_{\mu 4}|^2)^2 + 
|U_{\mu 4}|^4 e^{-\Gamma_4 L} \,.
\eeq
In the relevant $L/E_\nu$ range the most stringent bound on $\nu_\mu$
disappearance comes from the CDHS experiment~\cite{Dydak:1983zq}, where
the number of $\nu_\mu$ events in a near and far detector are
compared. Therefore this experiment is only sensitive to the $L$ 
dependent term in Eq.~(\ref{eq:mu-surv}). Since this term enters only
via $|U_{\mu 4}|^4$ the sensitivity of CDHS to $U_{\mu 4}$ is rather
poor in the decay scenario. Eq.~(\ref{eq:mu-surv}) has to be compared to
\beq
P_{\nu_\mu \to \nu_\mu}^\mathrm{SBL,osc} = 1 -  
4 |U_{\mu 4}|^2 (1 - |U_{\mu 4}|^2) 
\sin^2\frac{\Delta m^2_{41} L}{4E}
\eeq
for oscillations in a (3+1) four-neutrino scheme. In this case 
$L$-dependent effects enter already at order $|U_{\mu 4}|^2$,
which leads
to significantly stronger constraints than in the case of decay.
Technical details of our CDHS data analysis can be found in
Ref.~\cite{Grimus:2001mn}.

As shown in Ref.~\cite{atm4nu}, also atmospheric neutrino data provide
a non-trivial bound on the mixing of $\nu_\mu$ with heavy mass states,
i.e. on $|U_{\mu 4}|^2$.  Following Ref.~\cite{atm4nu}, in the case
of (3+1) oscillations, the survival probability of atmospheric
$\nu_\mu$ is given to a good approximation by
\beq\label{eq:atm-osc}
P_{\nu_\mu \to \nu_\mu}^\mathrm{ATM,osc} 
\approx P_l + |U_{\mu 4}|^4 \,,
\eeq
where $P_l$ is an effective survival probability involving only
oscillations of the light mass states $m_l$, which is obtained by
numerically solving the evolution equation within the Earth matter. Note
that the parameter $|U_{\mu 4}|^2$ enters also in $P_l$. The recent
update of a four-neutrino analysis of atmospheric~\cite{sk-atm},
and K2K~\cite{Aliu:2004sq} neutrino data performed in
Ref.~\cite{Maltoni:2004ei} gives the bound
\beq\label{eq:Umu4-bound}
|U_{\mu 4}|^2 \le 0.065\qquad (99\%~\mathrm{C.L.}) \,,
\eeq
which holds for $m_4 \gg \sqrt{\Delta m^2_\mathrm{ATM}} \sim 0.05$~eV,
where $\Delta m^2_\mathrm{ATM}$ is the mass-squared difference
resposible for atmospheric neutrino oscillations. In the decay
scenario the survival probability of atmospheric neutrinos is modified
to
\beq\label{eq:atm-dec}
P_{\nu_\mu \to \nu_\mu}^\mathrm{ATM,dec} 
\approx 
P_l + |U_{\mu 4}|^4 e^{-\Gamma_4 L}
\approx
P_l \,,
\eeq
where in the last step we have used the fact that for decay rates relevant to
LSND, we have $\Gamma_4 L \gg 1$ for atmospheric baseline and energy
ranges. Comparing Eqs.~(\ref{eq:atm-osc}) and (\ref{eq:atm-dec}) we
find that neglecting the small term $|U_{\mu 4}|^4$ the constraint
obtained for oscillations applies also in the case of decay. Note that
the decay will give a small contribution (suppressed by $|U_{\mu
4}|^2$) to the $P_{\nu^r_\mu \to \nu^s_e}$ transition probability for
atmospheric neutrinos, which potentially affects $e$-like events. In
addition, in general one has to take into account also oscillations of
the decay products with $\Delta m^2_{ll'}$ for atmospheric neutrinos,
and some of the $\nu^r_e$ produced in the decay will oscillate back to
$\nu_\mu^r$. However, under the assumption $R_e \approx 1,
R_{\mu,\tau} \approx 0$ for the branching ratios, oscillations of the
decay products with $\Delta m^2_\mathrm{ATM}$ will be doubly-suppressed
by $|U_{\mu 4} U_{e 3}|^2$. In the following decay analysis we will
neglect such subleading effects in atmospheric neutrinos, and we will
use as an approximation the $\chi^2(|U_{\mu 4}|^2)$ from Fig.~19 of
Ref.~\cite{Maltoni:2004ei}, which has been obtained for
oscillations. A detailed investigation of atmospheric neutrino data
within the decay scenario is beyond the scope of the present work.

\begin{figure}[t]
\centering
\includegraphics[width=0.95\textwidth]{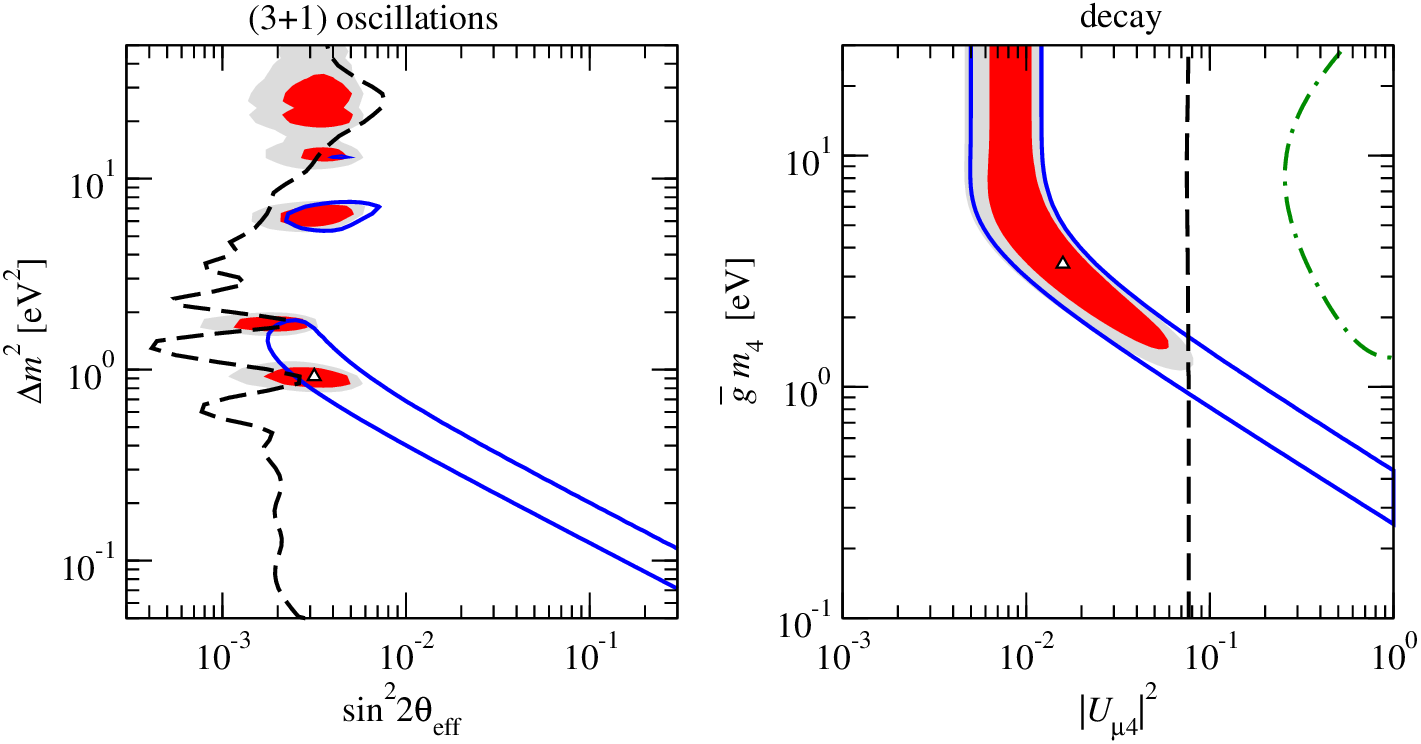}
  \mycaption{Allowed regions for LSND+KARMEN (solid) and SBL
  disappearance+atmospheric neutrino experiments (dashed) at 99\%~CL,
  and the combination of these data (shaded regions) at 90\% and
  99\%~CL. The left panel corresponds to neutrino oscillations in the
  (3+1) mass scheme and the right panel to the decay scenario
  presented in this work. The dash-dotted curve in the right panel
  shows the 99\%~CL constraint from CDHS.}
\label{fig:app-vs-disapp}
\end{figure}

Now we turn to the global analysis of all SBL data, and we investigate
the compatibility of LSND and null-result experiments within the decay
scenario. In Fig.~\ref{fig:app-vs-disapp} we show the allowed regions
from the appearance experiments LSND+KARMEN compared with the bound
implied from disappearance experiments. In this bound we include data
from the $\bar\nu_e \to \bar\nu_e$ reactor experiments
Bugey~\cite{Declais:1994su}, CHOOZ~\cite{Apollonio:2002gd}, and Palo
Verde~\cite{Boehm:2001ik}, as well as $\nu_\mu$ data from
CDHS~\cite{Dydak:1983zq}, atmospheric neutrinos~\cite{sk-atm}, and the
K2K long-baseline experiment~\cite{Aliu:2004sq}, as described above.

In the left panel of Fig.~\ref{fig:app-vs-disapp} we reproduce the
well-known result (see e.g. Refs.~\cite{Grimus:2001mn,atm4nu}) that,
for (3+1) mass schemes, the effective oscillation amplitude $\sin^2
2\theta_\mathrm{eff} = 4|U_{e 4}|^2|U_{\mu 4}|^2$ is tightly
constrained from the disappearance experiments by the quadratic
appearance of the small parameters $|U_{e 4}|^2$ and $|U_{\mu4}|^2$.
In contrast, in case of decay, appearance and disappearance experiments
are in perfect agreement, as is clear from the right panel of
Fig.~\ref{fig:app-vs-disapp}; there is a large overlap of the allowed 
regions for both data sets. The only relevant constraint from
disappearance data comes from the bound from atmospheric neutrinos
shown in Eq.~(\ref{eq:Umu4-bound}). As mentioned above no constraint
arises from reactor experiments, and in agreement with the discussion
related to Eq.~(\ref{eq:mu-surv}) one can see in
Fig.~\ref{fig:app-vs-disapp} that the constraint from CDHS is too weak 
to contribute significantly within the decay framework. The shaded
regions in the figure show the allowed regions obtained from combining
all data, where the total number data points is
\beq\label{eq:data}
11_\mathrm{(LSND)} +
9_\mathrm{(KARMEN)} + 15_\mathrm{(CDHS)} + 60_\mathrm{(Bugey)} +
1_\mathrm{(CHOOZ)} + 1_\mathrm{(Palo\,Verde)} + 1_\mathrm{(ATM)} =
98 \,.
\eeq
The global analysis gives the following best fit parameters for the
decay scenario
\beq
|U_{\mu 4}|^2 = 0.016 \,,\quad
\bar g \, m_4 = 3.4\,\mathrm{eV}\,.
\eeq
Hence, for neutrino masses in the range 100~keV coupling constants of
order $10^{-5}$ are sufficient to make the decay fast enough. Let us
quantify the quality of the fit further using three different
statistical tests.
\begin{enumerate}
\item
First we compare the fit for oscillations and decay by using the
absolute values of the global $\chi^2$ minimum. We find
\beq
\chi^2_\mathrm{min,osc} = 96.9 \,,\qquad
\chi^2_\mathrm{min,dec} = 88.3 \,.
\eeq
Although the GOF implied by these numbers is good in both cases, thanks to
the large number of data points (see Eq.~(\ref{eq:data})), the
$\Delta\chi^2 = 8.6$ indicates that decay provides a significantly better
description of the global data.
\item
Next we use the PG~\cite{Maltoni:2003cu} to test the compatibility of
LSND with the rest of the data. Applying Eq.~(\ref{eq:PG}) to this
case we find
\beq
\begin{array}{ll@{\quad}l}
\mbox{osc:} & \chi^2_\mathrm{PG} = 21.8\,, & \mathrm{PG} = 1.8\times
10^{-3}\,\% \\ 
\mbox{dec:} &\chi^2_\mathrm{PG} = 6.2  \,, & \mathrm{PG} = 4.6\,\%
\end{array}
\qquad
\mbox{(LSND vs rest)}
\eeq
where the $\chi^2$ values have been evaluated for 2~dof corresponding
to the two parameters in common to the two data sets. The PG numbers
show that for oscillations there is a severe disagreement between LSND
and all the other experiments, whereas the PG is acceptable within the
decay framework. The reason for the rather small PG even for decay
comes from the slight conflict between LSND and KARMEN, which is
present in any of the scenarios. Note that for five-neutrino
oscillations in a (3+2) mass scheme, a similar analysis performed in
Ref.~\cite{Sorel:2003hf} yielded PG$_{(3+2)} = 2.1\%$,
slightly worse than the one for decay.
\item
The better agreement of the data for decay becomes even more
transparent if we test the compatibility of appearance and
disappearance experiments, i.e. similar to
Fig.~\ref{fig:app-vs-disapp}, we divide the global data into 
LSND+KARMEN and all the rest. Then the PG analysis gives
\beq
\begin{array}{ll@{\quad}l}
\mbox{osc:} & \chi^2_\mathrm{PG} = 16.6\,, & \mathrm{PG} = 2.5\times
10^{-2}\,\% \\ 
\mbox{dec:} &\chi^2_\mathrm{PG} = 1.2  \,, & \mathrm{PG} = 55\,\%
\end{array}
\qquad
\mbox{(app.\ vs disapp.)}
\eeq
In this analysis the conflict between LSND and KARMEN is removed,
since they are added up to one single data set. Therefore, the above
numbers confirm our conclusion from Fig.~\ref{fig:app-vs-disapp} that
appearance and disappearance experiments are in perfect agreement in
the decay scenario.
\end{enumerate}

Before concluding this Section, let us mention again that we have
performed the analysis by setting $U_{e4} = 0$ and $R_e = 1,
R_\mu=0$. This reduces the number of parameters, and hence simplifies
the analysis considerably. Although we do not expect a significant
change of our results by relaxing these assumptions, we stress that on
general grounds this can only {\it improve} the fit of the decay
scenario, since more parameters become available to describe the data.

\section{Predictions for MiniBooNE and other future experiments}
\label{sec:mini} 

\subsection{MiniBooNE}

A critical test of the LSND signal will come from the MiniBooNE
experiment~\cite{miniboone}, which is currently taking data. This
experiment looks for $\nu_e$ appearance in a beam of $\nu_\mu$
neutrinos with a mean energy of $\sim 700$~MeV at a baseline of
$L\simeq 540$~m. In the minimal decay framework discussed in the
previous Sections we predict for MiniBooNE the appearance of $\nu_e$
events with a probability of the same order as in LSND, similar to the
case of oscillations. Since the MiniBooNE detector cannot distinguish
between neutrinos and antineutrinos, the $\bar\nu_e$ produced in the
helicity-flipping decay $n_4 \to \bar\nu_e \phi$ will also contribute to the
signal. We calculate the differential number of events as
\beq\label{eq:Nminib}
\frac{dN}{dE_{\nu_e}} \propto 
\int_{E_{\nu_e}}^\infty
dE_{\nu_\mu} \, \phi(E_{\nu_\mu}) \,
\left[
\frac{dP_{\nu_\mu \to \nu_e}(E_{\nu_\mu})}{dE_{\nu_e}}
\, \sigma(E_{\nu_e}) 
+
\frac{dP_{\nu_\mu \to \bar\nu_e}(E_{\nu_\mu})}{dE_{\bar\nu_e}}
\, \bar\sigma(E_{\nu_e}) 
\right]\,,
\eeq
where $\sigma$ ($\bar\sigma$) is the total charged-current cross
section for neutrinos (antineutrinos), and $\phi(E_{\nu_\mu})$ is the
initial flux of $\nu_\mu$, which we extract from Fig.~3 of
Ref.~\cite{miniboone}. 

\begin{figure}[t]
\centering
\includegraphics[width=0.6\textwidth]{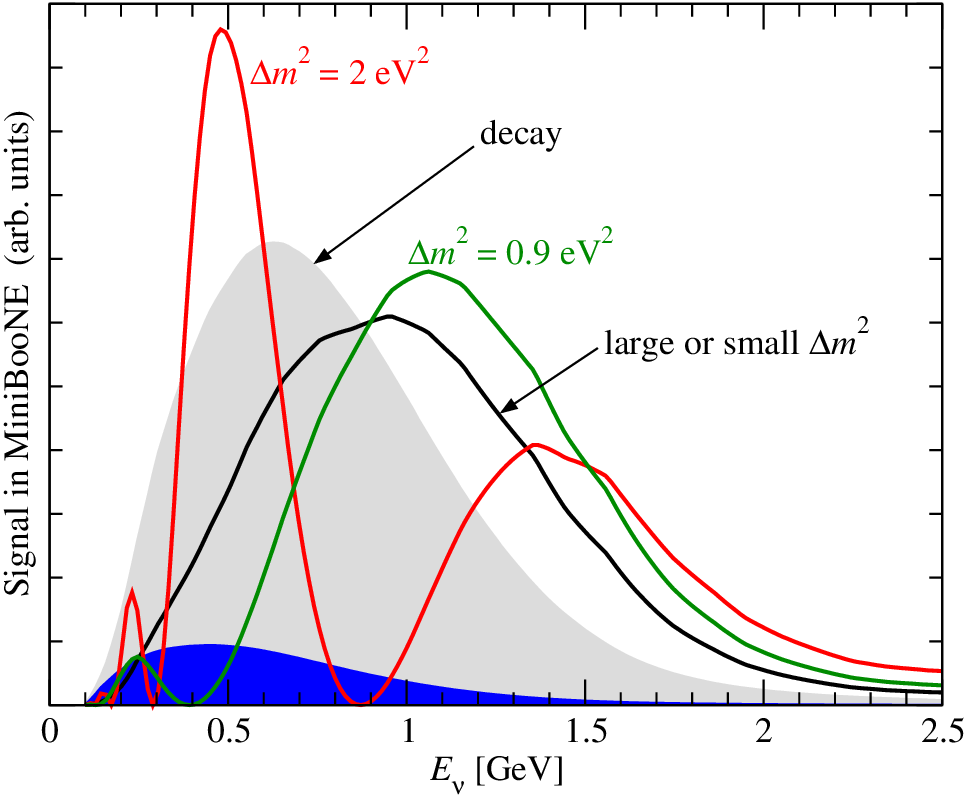}
  \mycaption{Energy spectrum predicted for the $\nu_e$ appearance
  signal in MiniBooNE for decay with $\bar g m_4 = 3.4$~eV (shaded
  region) and for various values of $\Delta m^2$ in the case of
  oscillations (solid curves). The blue/dark-shaded region shows the
  contribution of $\bar\nu_e$ from the helicity changing decay.}
\label{fig:miniboone}
\end{figure}

The spectral shape predicted by Eq.~(\ref{eq:Nminib}) is shown in
Fig.~\ref{fig:miniboone}. Although the heavy neutrino mass state
decays with equal probability into $\nu_e$ and $\bar\nu_e$, the
antineutrinos give only a small contribution to the total signal. The
reason is that, according to Eq.~(\ref{eq:W}), most of the decay
products from the helicity-flipping decay will have low energies, for
which the detection cross section is small. Moreover, the detection
cross section for antineutrinos is roughly a factor two smaller than
the one for neutrinos. Note that the spectral distortion implied by
$W_{rs}(E_{\nu_\alpha},E_{\nu_\beta})$ will be less pronounced in
MiniBooNE than in LSND because of the very different initial spectra.
In the LSND experiment, the spectrum rises monotonically up to a
maximum energy, whereas the MiniBooNE spectrum decreases with energy.
Therefore, degrading the energy of the decay products has a smaller
impact on the final spectrum in the case of MiniBooNE.

In Fig.~\ref{fig:miniboone} we show also the predicted spectral shape
in the case of oscillations for several values of $\Delta m^2$. One
observes that in principle the decay predicts a specific spectral shape of
the signal, whereas the actual spectrum from oscillations depends
strongly on $\Delta m^2$. Whether it will be possible to distinguish
between the two models in the case of a positive signal in MiniBooNE
depends on the available statistics, as well as on experimental factors
such as the energy resolution for $E_\nu$, which are not taken into
account in Fig.~\ref{fig:miniboone}.

Another method to discriminate between decay and oscillations is
provided by the disappearance channel in MiniBooNE. In the case of
(3+1), as well as (3+2) oscillations sizeable $\nu_\mu$ disappearance
is predicted, which should be observable in
MiniBooNE~\cite{sorel-thesis}. In contrast, the signal is expected to 
be very small in the decay scenario. The disappearance search in
MiniBooNE is limited to a shape analysis of the energy spectrum, since
the normalisation of the total number of events suffers from large
uncertainties~\cite{sorel-thesis}. Therefore, the signal for decay is
suppressed for the same reason as discussed above in relation with
Eq.~(\ref{eq:mu-surv}) and the CDHS experiment.

\subsection{Experiments looking for a non-zero value of $U_{e3}$}

In general the decay will contribute to the signal in experiments such
as T2K~\cite{t2k} or NO$\nu$A~\cite{Ayres:2004js} looking for a
non-zero value of $U_{e3}$ by exploring the chanel $\nu_\mu\to\nu_e$.
However, the appearance probability would be proportional to $|U_{\mu
4}|^2 \sim 0.01$, and thus at the edge of the sensitivity for those
experiments. Therefore, it appears to be rather challenging first to
observe an effect of our decay scenario, and second to disentangle it
from oscillations induced by $U_{e3}$. Much as in MiniBooNE, the
signal from decay will have a characteristic spectral shape, as
implied by the spectrum of the decay products. Moreover, in detectors
capable of distinguishing neutrinos from antineutrinos the appearance
of ``wrong-helicity'' neutrinos with low energies would be a generic
prediction of the decay scenario. Nevertheless, this is also very
challenging, because charge discrimination for electrons is a difficult
experimental task. In this respect, an interesting possibility to
explore the decay scenario might be to look for the appearance of
$\bar\nu_e$ from the $\nu_\mu$ beam of T2K in the KamLAND detector.
Let us note that also the Miner$\nu$a experiment~\cite{minerva}, which
will be placed $\sim 1$~km away from the target within the NuMI
neutrino beam, or the K2K/T2K near detectors could be suitable places
to look at these effects, although these beams might suffer from a
large intrinsic $\nu_e$ ($\bar\nu_e$) background.

Another manifestation of the decay model could be the observation of
$\nu_\mu\to\nu_e$ appearance in accelerator experiments such as T2K or
NO$\nu$A, but no corresponding signal for $\bar\nu_e$ disappearance in
reactor experiments such as Double-Chooz~\cite{Ardellier:2004ui}. This
would imply very small values of $U_{e3}$, such that $\bar\nu_e$
disappearance is suppressed, whereas the $\nu_\mu\to\nu_e$ appearance
signal is dominated by the decay. Note, however, that a positive
signal for $\bar\nu_e$ disappearance in reactor experiments cannot
exclude the decay scenario, since our model is perfectly compatible
with a sizeable value of $U_{e3}$.

In summary, we stress that a direct measurement of the decay effects
is quite challenging for present and near-future experiments looking
for a non-zero value of $U_{e3}$, because of the rather small mixing,
$|U_{\mu4}|^2$, which is needed in order to accommodate the LSND
result. Therefore, if the LSND signal should be confirmed by MiniBooNE
a new experiment at a stopped pion neutrino source as proposed in
Ref.~\cite{SNS} could be an optimal experiment to confirm or exclude
our decay model.

\section{CP-violating decays and the signal in MiniBooNE}
\label{sec:CP}

In this Section we extend our model and consider the case of two heavy
neutrinos, i.e. $N=5$ massive neutrinos. We will show that in this
case it is possible to obtain an interference between oscillation and
decay amplitudes~\cite{LOW01}, which may lead to CP violation in the
decay. In this way the signal in MiniBooNE (running in the neutrino
mode) can be significantly suppressed with respect to the LSND signal
from antineutrinos. 

Using Eq.~(\ref{eq:P_diff}) for $N=5$ and performing the
integral over $L$, one finds 
\begin{eqnarray}
P_{\nu_\alpha\to\nu_\beta}^\mathrm{dec}
&\equiv&
  \int_0^L dL' \frac{dP_{\nu_\alpha\to\nu_\beta}^\mathrm{dec}}{dL'}
  \nonumber\\
&=&
  \sum_h \frac{1}{2} \, |U_{\alpha h}|^2 \, R_{h\beta} 
  \left(1 - e^{-\Gamma_h L}\right) \nonumber\\
&+&
  \overline{R}_\beta \, |U_{\alpha 4}||U_{\alpha 5}| \,\cos\gamma 
  \left[ \cos(\delta+\gamma) - e^{-\overline{\Gamma} L} 
  \cos(\delta+\gamma+\Gamma_\mathrm{osc}L) \right] ,
\label{eq:P_CP}
\end{eqnarray}
with the following definitions for the branching ratios
\beq
R_{h\beta} \equiv 2\frac{|g_{h\beta}|^2 A_h^2}{\Gamma_h}
\,,\quad
\overline{R}_\beta \equiv 2 \frac{|g_{4\beta}||g_{5\beta}| \, A_4 A_5}
{\overline{\Gamma}}
\,,\quad
\overline{\Gamma} \equiv \frac{\Gamma_4+\Gamma_5}{2} \,,
\eeq
and phases
\beq
\delta \equiv \mathrm{arg}(U_{\alpha 4}^*U_{\alpha 5} \,
g_{4\beta}g_{5\beta}^*) 
\,,\quad 
\tan\gamma \equiv \frac{\Gamma_\mathrm{osc}}{\overline{\Gamma}}
\,,\quad
\Gamma_\mathrm{osc} \equiv \frac{\Delta m^2_{54}}{2 E_{\nu_\alpha}} \,.
\eeq
The first term in Eq.~(\ref{eq:P_CP}) describes the incoherent decay
of the two heavy mass states in analogy to Eq.~(\ref{eq:P2}), whereas
the second term results from the interference of the amplitudes for
decay and oscillations of the two heavy neutrino mass states $n_4$ and
$n_5$. The phase $\delta$ is the analogue of the Dirac phase leading
to CP violation in the standard three neutrino oscillation framework.
In fact, in the case of antineutrinos, $U_{\alpha h}$ and $g_{h\beta}$
have to be replaced by $U_{\alpha h}^*$ and $g_{h\beta}^*$, which
implies $\delta \to -\delta$, and we obtain the CP asymmetry:
\begin{eqnarray}
\Delta P 
&\equiv& 
  P_{\bar\nu_\alpha \to \bar\nu_\beta}^\mathrm{dec} -
  P_{\nu_\alpha\to\nu_\beta}^\mathrm{dec} \nonumber\\
&=&
  2 \overline{R}_\beta \, |U_{\alpha 4}||U_{\alpha 5}| \,\cos\gamma \,
  \sin\delta
  \left[ \sin\gamma - e^{-\overline{\Gamma} L} 
  \sin(\gamma+\Gamma_\mathrm{osc}L) \right] \,.
\label{eq:assym}
\end{eqnarray}
One observes that necessary conditions to obtain CP violation are
$\delta \neq 0,\pi$ and $\tan\gamma \sim 1$, i.e. $\overline{\Gamma}
\sim \Gamma_\mathrm{osc}$ or $\bar g_h^2 m_h^2 / 8\pi \sim \Delta
m^2_{54}$. Using $\bar g_h m_h \sim$ few eV, the last condition shows
that the heavy neutrinos ($m_h >$ keV) have to be highly degenerate.

\begin{figure}[t]
\centering 
   \includegraphics[width=0.6\textwidth]{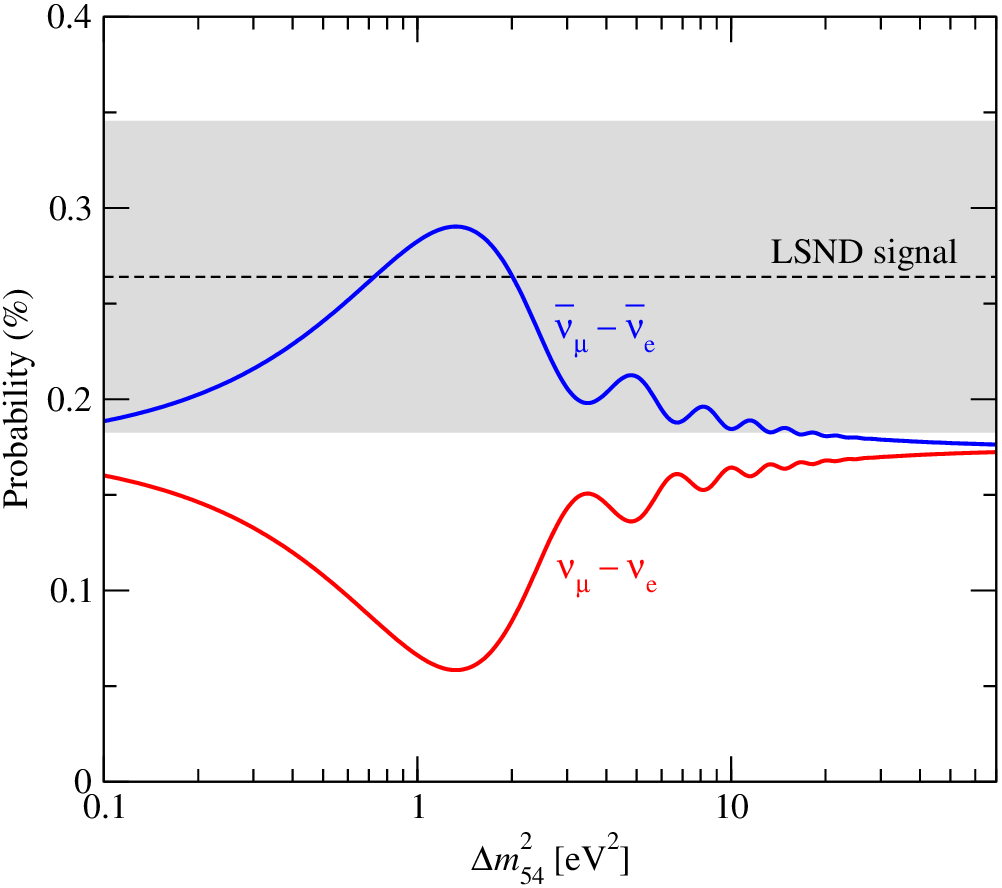}
   \mycaption{The transition probabilities $P_{\bar\nu_\mu \to
   \bar\nu_e}^\mathrm{dec}$ and $P_{\nu_\mu\to\nu_e}^\mathrm{dec}$
   according to Eq.~(\ref{eq:P_CP}). The shaded region corresponds to
   the $1\sigma$ range for $P_{\bar\nu_\mu \to \bar\nu_e}$ observed in
   LSND. The chosen parameter values are $L/E_\nu \simeq 0.75$~m/MeV,
   $|U_{\mu 4}|^2 = |U_{\mu 5}|^2 = 0.003$,
   $R_{4e}=R_{5e}=\overline{R}_e=1$, $\Gamma_4 = \Gamma_5 =
   \overline{\Gamma}$ with $\bar g_h m_h = 3.4$~eV, and $\delta =
   \pi/2$. }
   \label{fig:cp}
\end{figure}

In Fig.~\ref{fig:cp} we show a numerical example for the
$P_{\bar\nu_\mu \to \bar\nu_e}^\mathrm{dec}$ and
$P_{\nu_\mu\to\nu_e}^\mathrm{dec}$ transition probabilities for an
$L/E_\nu$ value relevant to LSND and MiniBooNE. Clearly, in this
example the neutrino signal is strongly suppressed with respect to the
antineutrino signal for $\Delta m^2_{54} \sim 1$~eV$^2$. This result
confirms the previous observation that in order to obtain CP
violation, the two heavy neutrinos have to be very degenerate, with
masses in the keV range and a mass difference in the eV range. Whether
this mechanism allows indeed to accommodate the LSND signal with a
possible null-result of the MiniBooNE experiment in the neutrino mode
has to be investigated after the release of the MiniBooNE data by a
combined analysis of the two experiments within this framework. In
that case, MiniBooNE data taking with an antineutrino beam would be necessary
in order to test the CP-violating decay scenario.

\section{Constraints from laboratory, cosmology, and astrophysics}
\label{sec:cosmo}

Mixing between active and sterile neutrinos\footnote{See
Ref.~\cite{Cirelli:2004cz} for a recent analysis of various bounds
on sterile neutrino mixing.}, and couplings between active neutrinos
and a light scalar, have been extensively studied, both in laboratory
experiments and, for their implications in the evolution of the early
Universe and of supernovae. For the discussion of these constraints we
will focus on the minimal scenario necessary to explain the LSND
signal: we require only mixing of $\nu_\mu$ with the heavy mass
state, $|U_{\mu 4}|^2 \sim 0.01$, whereas we set $U_{e 4} = U_{\tau 4} =
0$. Furthermore, the assumption $R_e =1, R_{\mu,\tau} = 0$ for the
branching ratios of the decay implies that only two elements of the
(symmetric) coupling matrix in the flavour basis, $g_{\alpha\beta}$,
are non-zero: $g_{es} \simeq \bar g$ and $g_{e\mu} \simeq U_{\mu 4} \,
\bar g$.

First we note that in our model solar neutrino oscillations are
completely unaffected. Because of the assumption $U_{e4}=0$, no $n_h$
component can be produced in the Sun, and since the $\nu_l$ in our
scenario are stable, there is no decay of solar neutrinos.  For the
very same reason the decay has also no effect in the KamLAND
experiment, and these data are entirely explained by oscillations of
the light neutrinos. On the other hand, effects in atmospheric
neutrino experiments have been discussed in Section~\ref{sec:SBL}.

\subsection{Laboratory bounds}

In general, the mixing of a heavy neutrino leads to contributions to
the effective neutrino masses in neutrinoless double-beta
decay~\cite{bbdecay} and tritium beta decay~\cite{tritium}
experiments. Note, however, that in our scenario there will be no
effect of the heavy neutrino in such experiments because of our assumption
$U_{e4} = 0$. The coupling between $\nu_e$ and a light scalar would
induce double-beta decay with the emission of one or two scalars, with
a spectrum for the two electrons which is distinguishable from the one
of the two-neutrino double-beta decay. Relatively strong bounds for
single scalar emission~\cite{nemo2} apply only to the coupling
$g_{ee}$, which can be arbitrary small in our scenario. The couplings
$g_{e\alpha}$ for $\alpha \neq e$ contribute in principle to
double-beta decay with the emission of two scalars. However, in this
case the limits are very weak~\cite{nemo2}: $g < \mathcal{O}(1)$.

The decay of pions and kaons has been used to set bounds on the mixing
of heavy neutrinos~(see Ref.~\cite{KPS04} and references therein).  
If a massive neutrino with $m_h \gg m_{1,2,3}$ were produced in such
decays, the energy spectrum of the muon would present an additional
monochromatic line. No positive signal was found, leading to strong
constraints on the mixing for neutrino with masses $m_h \gtap
1$~MeV. Mixing of neutrinos with smaller
masses are compatible with these bounds. Furthermore, light scalar
emission was not observed in pion and kaon
decays~\cite{piondecays,GSR82}. The most stringent bounds are of order 
$g^2 < \mathrm{few} \times 10^{-5}$, much too weak to constrain our
model.

\subsection{Supernova bounds}

To estimate the effect of our model in thermal environments such as in
a supernova and the early Universe let us compare the rate of
processes induced by the Lagrangian Eq.~(\ref{eq:Ldec}) to the one of
weak processes of the type $\nu e \to \nu e$.  From dimensional
considerations, it follows that
$\sigma_\mathrm{weak} \sim G_F^2 E^2$, where
$E$ is the typical energy of the involved particles. Using
$\sigma(2\leftrightarrow 1) \sim \bar g^2 m_h^2/ E^4$ and
$\sigma(2\leftrightarrow 2) \sim \bar g^4 / E^2$, where
$2\leftrightarrow 1$ denotes processes like $\nu_l\phi \leftrightarrow
n_h$, $\nu_l n_h \to \phi$, and $2\leftrightarrow 2$
indicates processes like $\nu_l\nu_l \leftrightarrow\phi\phi$,
$\nu_l\nu_l \leftrightarrow n_h n_h$, we find\footnote{For a
careful analysis of these processes for active neutrinos
see~Ref.~\cite{Farzan02}.}
\begin{eqnarray}
\frac{\sigma(2 \leftrightarrow 1)}{\sigma_\mathrm{weak}}
&\sim& 10^{10} 
\left(\frac{\bar g \, m_h}{1\,\mathrm{eV}}\right)^2
\left(\frac{E}{1\,\mathrm{MeV}}\right)^{-6} \,, 
\label{eq:sigma3}\\
\frac{\sigma(2 \leftrightarrow 2)}{\sigma_\mathrm{weak}}
&\sim& 10^2 
\left(\frac{\bar g}{10^{-5}}\right)^4
\left(\frac{E}{1\,\mathrm{MeV}}\right)^{-4} \,.
\label{eq:sigma4}
\end{eqnarray}
In addition to these reactions induced by the scalar coupling, heavy
mostly-sterile neutrinos are produced via oscillations due to their
mixing with the active ones.

If heavy neutrinos and light scalars were produced in the core of a
supernova and could escape freely, they would carry away a large amount
of energy, substantially modifying the supernova evolution. Together
with the observations from supernova SN1987A, this energy-loss
argument has been used to constrain sterile neutrino
mixing~\cite{SNsterile} as well as Majoron coupling, see e.g.
Refs.~\cite{Farzan02,SNmajoron}. In our scenario the mass of the heavy
neutrino is large compared with the matter potential within the
supernova core. Therefore, we can treat mixing, as well as the
reactions involving the scalar and $n_h$, as in vacuum. From
Eq.~(\ref{eq:sigma3}) it follows that for typical energies in the
supernova cores, $E\sim 10$~MeV, $2 \leftrightarrow 1$ reactions are
$10^4$ times faster than weak interactions. This implies that $n_h$
and $\phi$ are strongly coupled to the active neutrinos, and hence,
they are trapped within the neutrinosphere, avoiding any energy loss
due to particles escaping from the core. Let us add that there might
be additional effects, such as modification of lepton number
or neutrino flux distortions~\cite{GSR82,delep}. The analysis of such
effects requires detailed studies, which are beyond the scope of the
present work.

\subsection{Cosmological bounds}

Constraints on the number of relativistic degrees of freedom present
at the time of Big Bang Nucleosynthesis (BBN), usually parametrized by
the number of neutrino flavours $N_\nu$, are relevant when light
particles such as sterile neutrinos are
introduced~\cite{sterileBBN}. Analyses of cosmological data, including
light-element abundances lead to the limits $2.3\le N_\nu \le 3.0$
(95\%~CL)~\cite{DeltaN}. A recent analysis, that uses a new assessment
of the primordial $^4$He abundance and its uncertainty, has
substantially relaxed the bounds on $N_\nu\geq 3$: $\delta N_\nu \le
1.44$ (95\%~CL)~\cite{Cyburt:2004yc}, where $\delta N_\nu \equiv N_\nu
- 3$. The discrepancy between these two exemplary results illustrates
that such bounds have to be considered with care, because of the
systematical uncertainties in light-element abundances, and the
dependence on input priors.

In the early Universe, $n_h$ of our decay model are initially
generated by mixing with active neutrinos~\cite{barbieristerile}. As
soon as $n_h$ are produced, they are thermalized together with the
scalar due to the very fast $2\leftrightarrow 1$ reactions. Hence, one
additional neutrino and one scalar degree of freedom will be present
during BBN, leading to $N_\nu = 4.57$, which appears to be disfavoured
by the limits quoted above.\footnote{Note that also in four- or
five-neutrino oscillation schemes the sterile neutrinos are
thermalized within the standard BBN scenario~\cite{sterileBBN}.}
However, to draw reliable conclusions from BBN considerations a more
detailed analysis is required, since there are many effects which
might play a role. For example, in the presence of a small chemical
potential for $\nu_e$ the limit on extra relativistic degrees of
freedom becomes much weaker, $N_\nu \le 6.5$~\cite{DeltaN}, and hence
$N_\nu = 4.57$ is allowed. Moreover, a detailed simulation of BBN in
the presence of $\tau$-neutrino decay performed in
Ref.~\cite{Dolgov:1998st} showed that the effective $N_\nu$ depends in
a rather non-trivial way on model parameters, and values of $\delta
N_\nu > 0$ as well as $< 0$ may be obtained.
Let us mention also the possibility to avoid the thermalization of
$n_h$ and $\phi$ at BBN, by suppressing the initial production of
$n_h$ through mixing. This could be achieved e.g.\ through a
large lepton asymmetry of order $10^{-3}$~\cite{FV95}, a matter
potential induced by neutrino--Majoron interactions~\cite{BB01}, or by
assuming a low reheating temperature~\cite{GPRP}.  The
$2\leftrightarrow 2$ processes would not be efficient in producing and
thermalizing sterile neutrinos and scalars as far as they are slower
than the expansion rate at BBN. Couplings $g \ltap \rm{few} \times
10^{-6}$ guarantee that these scatterings are out of equilibrium for
$T \gtap 100$~keV (see Eq.~(\ref{eq:sigma4})).
 
As can be seen from Eqs.~(\ref{eq:sigma3}) and (\ref{eq:sigma4}) the
scalar interactions have a recoupling form, instead of the usual
decoupling form of weak interactions, i.e. their strength grows as
the Universe evolves, which may have some effects in later epochs of
the Universe. For instance, if a light scalar couples to neutrinos
before the recombination era, effects like a delayed matter--radiation
equality and enhanced damping of acoustic oscillations at higher $l$
can occur because the radiation energy density differs from that of
the standard scenario. Moreover, if one or more 
neutrinos are scattering during the eV era rather than free-streaming,
a uniform shift of the CMB peaks to larger $l$ would be
manifest~\cite{CMBsignal}.

Finally, we note that our scenario, with the scalar heavier than
the light neutrinos, does not give rise to a neutrinoless
Universe~\cite{BBD04}, which is disfavoured by structure formation
arguments~\cite{Hannestad04}.

\section{Summary and conclusions}
\label{sec:conclusions}

We have presented an explanation for the LSND evidence for $\bar\nu_e$
appearance based on the decay of a heavy sterile neutrino into a light
scalar particle and light neutrinos. We assume a small mixing of the
heavy neutrino mass eigenstate $n_4$ with $\nu_\mu$ of order $|U_{\mu
4}|^2 \sim 0.01$, such that a small $n_4$ component is contained in
the initial $\bar\nu_\mu$ beam produced in the LSND experiment. On the
way to the LSND detector the $n_4$ decays into the scalar and a
superposition of light neutrino mass eigenstates. If this final state
contains a large fraction of the $\bar\nu_e$ flavour, it will give
rise to the $\bar\nu_e$ appearance signal observed in LSND. Taking
into account the energy distribution of the decay products, one
obtains a characteristic prediction for the spectral shape of the LSND
signal, with more events at the lower part of the spectrum. Comparing
this prediction with the spectral data of LSND, we find that the fit
of the decay is slightly worse than for oscillations, although still
acceptable ($\chi^2_\mathrm{decay} = 10.8/9$~dof). A more detailed
investigation of the LSND spectral data might allow to distinguish
between decay and oscillations.

We have performed a global fit to all relevant data, including
null-result short-baseline reactor and accelerator oscillation
experiments. The agreement of the $\bar\nu_\mu\to\bar\nu_e$ appearance
experiments LSND and KARMEN in the decay scenario is at the same level
as in the case of oscillations. Much as the oscillation phase, also
the decay exponential depends on $L/E_\nu$, which allows us to reconcile
the two results due to the different baselines. However, unlike the
case of (3+1) four-neutrino schemes, our decay model is in complete
agreement with the constraints from disappearance results. Using the
so-called parameter goodness of fit (PG) to test the compatibility of
appearance (LSND, KARMEN) and disappearance (Bugey, CHOOZ, CDHS,
atmospheric neutrino) experiments, we find PG~$= 55\%$ for decay but
only PG~$=0.025\%$ for (3+1) oscillations. Testing the compatibility of
LSND and all the null-result experiments, we find PG~$= 4.6\%$ for
decay, which is slightly better than the PG~$=2.1\%$ obtained in
Ref.~\cite{Sorel:2003hf} for a (3+2) five-neutrino oscillation
scenario. 
In addition, we have shown that our model is consistent with
present laboratory bounds and have discussed implications for
supernova evolution, BBN and CMB observations.  

In the minimal version of the decay model we predict a signal in the
upcoming MiniBooNE experiment corresponding to a transition
probability of the same order as seen in LSND, and we have discussed
characteristic signatures of our scenario, which may allow to
distinguish it from oscillations in MiniBooNE or other future neutrino
experiments. Furthermore, we have shown that within an extension of
the minimal decay model it is possible to introduce CP violation in
the decay through an interference term between decay and oscillations
of two nearly degenerate heavy neutrinos, with masses in the
1~keV--1~MeV range and $\Delta m^2_{54} \sim 1$~eV$^2$. This can lead
to a suppression of the signal in MiniBooNE running in the neutrino
mode, and simultaneously accounting for the antineutrino signal in
LSND.  Hence, this model can only be excluded if MiniBooNE does not
find a positive signal either in the neutrino mode or running with
antineutrinos.

\subsection*{Acknowledgements}

We thank J.~Beacom, B.~Louis, S.~Pakvasa, S.~Pastor, S.T.~Petcov,
H.~Ray, R.~Scherrer, M.~Sorel, R.~Tomas and W.~Winter for useful
discussions. S.P.R.\ and S.P.\ are grateful to the Fermilab
Theoretical Astrophysics Group for kind hospitality during part of
this work. T.S.\ acknowledges support from the CERN Theory group for a
visit during which this work was initiated. S.P.R.\ is supported by
NASA Grant ATP02-0000-0151 and by the Spanish Grant FPA2002-00612 of
the MCT. T.S.\ is supported by a ``Marie Curie Intra-European
Fellowship within the 6th European Community Framework Programme''.



\begin{thebibliography}{99}

\bibitem{Aguilar:2001ty} 
  A.~Aguilar {\it et al.}  [LSND Coll.], 
  Phys.\ Rev.\ D {\bf 64}, 112007 (2001)
  [hep-ex/0104049].

\bibitem{karmen}
  B.~Armbruster {\it et al.}  [KARMEN Coll.],
  Phys.\ Rev.\ D {\bf 65}, 112001 (2002)
  [hep-ex/0203021].


\bibitem{sk-atm}
  Y.~Fukuda {\it et al.}  [Super-Kamiokande Coll.],
  Phys.\ Rev.\ Lett.\  {\bf 81}, 1562 (1998)
  [hep-ex/9807003];
%
  Y.~Ashie {\it et al.},  
  Phys.\ Rev.\ D {\bf 71}, 112005 (2005)
  [hep-ex/0501064].

\bibitem{dip}
  Y.~Ashie {\it et al.}  [Super-K Coll.],
  Phys.\ Rev.\ Lett.\  {\bf 93}, 101801 (2004)
  [hep-ex/0404034].


\bibitem{Aliu:2004sq}
  E.~Aliu {\it et al.}  [K2K Coll.],
  Phys.\ Rev.\ Lett.\  {\bf 94}, 081802 (2005)
  [hep-ex/0411038].


\bibitem{sno}
  Q.~R.~Ahmad {\it et al.}  [SNO Coll.],
  Phys.\ Rev.\ Lett.\  {\bf 89}, 011301 (2002)
  [nucl-ex/0204008];
%
  B.~Aharmim {\it et al.},
  nucl-ex/0502021.


\bibitem{Araki:2004mb}
  T.~Araki {\it et al.}  [KamLAND Coll.],
  Phys.\ Rev.\ Lett.\  {\bf 94}, 081801 (2005)
  [hep-ex/0406035].


\bibitem{sterile}
  J.~T.~Peltoniemi, D.~Tommasini and J.~W.~F.~Valle,
  Phys.\ Lett.\ B {\bf 298}, 383 (1993);
%
  J.~T.~Peltoniemi and J.~W.~F.~Valle,
  Nucl.\ Phys.\ B {\bf 406}, 409 (1993)
  [hep-ph/9302316];
%
  D.O. Caldwell and R.N. Mohapatra, 
  Phys.\ Rev.\ D {\bf 48}, 3259 (1993).
  

\bibitem{Maltoni:2002xd}
  M.~Maltoni {\it et al.},
  Nucl.\ Phys.\ B {\bf 643}, 321 (2002) 
  [hep-ph/0207157].


\bibitem{strumia}
  A.~Strumia, 
  Phys.\ Lett.\ B {\bf 539}, 91 (2002)
  [hep-ph/0201134].


\bibitem{Maltoni:2004ei}
  M.~Maltoni {\it et al.},
  New J.\ Phys.\  {\bf 6}, 122 (2004)
  [hep-ph/0405172].


\bibitem{Dydak:1983zq}
  F.~Dydak {\it et al.},
  Phys.\ Lett.\ B {\bf 134}, 281 (1984).


\bibitem{Declais:1994su}
  Y.~Declais {\it et al.},
  Nucl.\ Phys.\ B {\bf 434}, 503 (1995).


\bibitem{Astier:2003gs}
  P.~Astier {\it et al.}  [NOMAD Coll.],
  Phys.\ Lett.\ B {\bf 570}, 19 (2003)
  [hep-ex/0306037].


\bibitem{Ma:1999im}
  E.~Ma, G.~Rajasekaran and I.~Stancu,
  Phys.\ Rev.\ D {\bf 61}, 071302 (2000) [hep-ph/9908489];
%
  E.~Ma and G.~Rajasekaran,
  Phys.\ Rev.\ D {\bf 64}, 117303 (2001) [hep-ph/0107203].


\bibitem{cpt}
  H.~Murayama and T.~Yanagida,
  Phys.\ Lett.\ B {\bf 520}, 263 (2001)
  [hep-ph/0010178];
%
  G.~Barenboim, L.~Borissov and J.~Lykken, 
  hep-ph/0212116;
%
  M.~C.~Gonzalez-Garcia, M.~Maltoni and T.~Schwetz,
  Phys.\ Rev.\ D {\bf 68}, 053007 (2003)
  [hep-ph/0306226].


\bibitem{LNV}
  K.~S.~Babu and S.~Pakvasa,
  hep-ph/0204236;
%
  B.~Armbruster {\it et al.} [KARMEN Coll.],
  Phys.\ Rev.\ Lett.\  {\bf 90}, 181804 (2003)
  [hep-ex/0302017];
%
  A.~Gaponenko {\it et al.}  [TWIST Coll.],
  Phys.\ Rev.\ D {\bf 71}, 071101 (2005)
  [hep-ex/0410045].

\bibitem{Sorel:2003hf}
  M.~Sorel, J.~M.~Conrad and M.~Shaevitz,
  Phys.\ Rev.\ D {\bf 70}, 073004 (2004)
  [hep-ph/0305255].


\bibitem{4nu-cpt}
  V.~Barger, D.~Marfatia and K.~Whisnant,
  Phys.\ Lett.\ B {\bf 576}, 303 (2003)
  [hep-ph/0308299].


\bibitem{GPRP}
  G.~Gelmini, S.~Palomares-Ruiz and S.~Pascoli,
  Phys.\ Rev.\ Lett.\  {\bf 93}, 081302 (2004)
  [astro-ph/0403323].


\bibitem{Barenboim:2004wu}
  G.~Barenboim and N.~E.~Mavromatos,
  JHEP {\bf 0501}, 034 (2005)
  [hep-ph/0404014].


\bibitem{MVN}
  D.~B.~Kaplan, A.~E.~Nelson and N.~Weiner,
  Phys.\ Rev.\ Lett.\  {\bf 93}, 091801 (2004) 
  [hep-ph/0401099];
%
  K.~M.~Zurek,
  JHEP {\bf 0410}, 058 (2004)
  [hep-ph/0405141].
 

\bibitem{Pas:2005rb}
  H.~Paes, S.~Pakvasa and T.~J.~Weiler,
  hep-ph/0504096.


\bibitem{miniboone}
  J.~Monroe [MiniBooNE Coll.],
  hep-ex/0406048.

\bibitem{majoronmodels}
  Y.~Chikashige, R.~N.~Mohapatra and R.~D.~Peccei,
  Phys.\ Rev.\ Lett.\  {\bf 45}, 1926 (1980) and
%
  Phys.\ Lett.\ B {\bf 98}, 265 (1981);
%
  G.~B.~Gelmini and M.~Roncadelli,
  Phys.\ Lett.\ B {\bf 99}, 411 (1981);
%
  H.~M.~Georgi, S.~L.~Glashow and S.~Nussinov,
  Nucl.\ Phys.\ B {\bf 193}, 297 (1981);
%
  G.~B.~Gelmini and J.~W.~F.~Valle,
  Phys.\ Lett.\ B {\bf 142}, 181 (1984);
%
  A.~Santamaria and J.~W.~F.~Valle,
  Phys.\ Rev.\ Lett.\  {\bf 60}, 397 (1988);
%
  S.~Bertolini and A.~Santamaria,
  Nucl.\ Phys.\ B {\bf 310}, 714 (1988).


\bibitem{nudecay}
  A.~Acker, A.~Joshipura and S.~Pakvasa,
  Phys.\ Lett.\ B {\bf 285}, 371 (1992).


\bibitem{atmosdecay}
  V.~D.~Barger {\it et al.},
  Phys.\ Rev.\ Lett.\  {\bf 82}, 2640 (1999)
  [astro-ph/9810121];
%
  P.~Lipari and M.~Lusignoli,
  Phys.\ Rev.\ D {\bf 60}, 013003 (1999)
  [hep-ph/9901350];
%
  G.~L.~Fogli {\it et al.}, 
  Phys.\ Rev.\ D {\bf 59}, 117303 (1999)
  [hep-ph/9902267];
%
  S.~Choubey and S.~Goswami,
  Astropart.\ Phys.\  {\bf 14}, 67 (2000)
  [hep-ph/9904257];
%
  V.~D.~Barger {\it et al.}, 
  Phys.\ Lett.\ B {\bf 462}, 109 (1999)
  [hep-ph/9907421].


\bibitem{solardecay}
  J.~N.~Bahcall, N.~Cabibbo and A.~Yahil,
  Phys.\ Rev.\ Lett.\  {\bf 28}, 316 (1972);
%
  S.~Pakvasa and K.~Tennakone,
  Phys.\ Rev.\ Lett.\  {\bf 28}, 1415 (1972);
%
  J.~N.~Bahcall {\it et al.}, 
  Phys.\ Lett.\ B {\bf 181}, 369 (1986);
%
  S.~Choubey, S.~Goswami and D.~Majumdar,
  Phys.\ Lett.\ B {\bf 484}, 73 (2000)
  [hep-ph/0004193];
%
  A.~Bandyopadhyay, S.~Choubey and S.~Goswami,
  Phys.\ Rev.\ D {\bf 63}, 113019 (2001)  
  [hep-ph/0101273];
%
  J.~F.~Beacom and N.~F.~Bell,
  Phys.\ Rev.\ D {\bf 65}, 113009 (2002)
  [hep-ph/0204111].


\bibitem{karmen1}
  B.~Armbruster {\it et al.} [KARMEN Coll.],
  Phys.\ Lett.\ B {\bf 348}, 19 (1995).


\bibitem{karmen2}
  K.~Eitel  [KARMEN Coll.],
  Nucl.\ Phys.\ Proc.\ Suppl.\  {\bf 91}, 191 (2000)
  [hep-ex/0008002].


\bibitem{boundsdecay}
  R.~E.~Shrock,
  Phys.\ Rev.\ D {\bf 24}, 1275 (1981);
%
  N.~De Leener-Rosier {\it et al.},
  Phys.\ Rev.\ D {\bf 43}, 3611 (1991);
%
  M.~Daum {\it et al.},
  Phys.\ Lett.\ B {\bf 361}, 179 (1995);
%
  J.~A.~Formaggio {\it et al.}  [NuTeV Coll.],
  Phys.\ Rev.\ Lett.\  {\bf 84}, 4043 (2000)
  [hep-ex/9912062];
%
  M.~Daum {\it et al.},
  Phys.\ Rev.\ Lett.\  {\bf 85}, 1815 (2000)
  [hep-ex/0008014].


\bibitem{KPS04}
  A.~Kusenko, S.~Pascoli and D.~Semikoz,
  hep-ph/0405198.


\bibitem{Farzan02}
  Y.~Farzan,
  Phys.\ Rev.\ D {\bf 67}, 073015 (2003)
  [hep-ph/0211375].


\bibitem{Kim:1990km}
  C.~W.~Kim and W.~P.~Lam,
  Mod.\ Phys.\ Lett.\ A {\bf 5}, 297 (1990);
%
  C.~Giunti {\it et al.},
  Phys.\ Rev.\ D {\bf 45} (1992) 1557.


\bibitem{LOW01}
  M.~Lindner, T.~Ohlsson and W.~Winter,
  Nucl.\ Phys.\ B {\bf 607}, 326 (2001)  
  [hep-ph/0103170] and
  {\it ibid.} B {\bf 622}, 429 (2002) 
  [astro-ph/0105309].


\bibitem{Church:2002tc}
  E.~D.~Church {\it et al.},
  Phys.\ Rev.\ D {\bf 66}, 013001 (2002)
  [hep-ex/0203023].


\bibitem{Maltoni:2003cu}
  M.~Maltoni and T.~Schwetz,
  Phys.\ Rev.\ D {\bf 68}, 033020 (2003)
  [hep-ph/0304176].


\bibitem{Grimus:2001mn}
  W.~Grimus and T.~Schwetz,
  Eur.\ Phys.\ J.\ C {\bf 20}, 1 (2001) 
  [hep-ph/0102252].


\bibitem{atm4nu}
  S.~M.~Bilenky {\it et al.}, 
  Phys.\ Rev.\ D {\bf 60}, 073007 (1999)
  [hep-ph/9903454];
%
  M.~Maltoni, T.~Schwetz and J.~W.~F.~Valle,
  Phys.\ Lett.\ B {\bf 518}, 252 (2001)
  [hep-ph/0107150] and
%
  Phys.\ Rev.\ D {\bf 65}, 093004 (2002) 
  [hep-ph/0112103].


\bibitem{Apollonio:2002gd}
  M.~Apollonio {\it et al.},
  Eur.\ Phys.\ J.\ C {\bf 27}, 331 (2003)
  [hep-ex/0301017].


\bibitem{Boehm:2001ik}
  F.~Boehm {\it et al.},
  Phys.\ Rev.\ D {\bf 64}, 112001 (2001)
  [hep-ex/0107009].


\bibitem{sorel-thesis}
  M.~Sorel, PhD thesis, FERMILAB-THESIS-2005-07, available at \\ 
  http://library.fnal.gov/archive/thesis/fermilab-thesis-2005-07.shtml


\bibitem{t2k}
  Y.~Itow {\it et al.},
  hep-ex/0106019;  
%
  T. Kobayashi, talk at NNN05, 7--9 April 2005, Aussois, Savoie,
  France, http://nnn05.in2p3.fr/schedule.html


\bibitem{Ayres:2004js}
  D.~S.~Ayres {\it et al.} [NOvA Coll.],
  hep-ex/0503053.


\bibitem{minerva}
  D.~Drakoulakos {\it et al.}  [Minerva Coll.],
  hep-ex/0405002.


\bibitem{Ardellier:2004ui} 
  F.~Ardellier {\it et al.}, 
  hep-ex/0405032;
%
  K.~Anderson {\it et al.},
  hep-ex/0402041;
%
  P.~Huber {\it et al.}, 
  Nucl.\ Phys.\ B {\bf 665}, 487 (2003)
  [hep-ph/0303232].


\bibitem{SNS}
  G.~T.~Garvey {\it et al.},
  hep-ph/0501013.



\bibitem{Cirelli:2004cz}
  M.~Cirelli {\it et al.},
  Nucl.\ Phys.\ B {\bf 708}, 215 (2005)
  [hep-ph/0403158].


\bibitem{bbdecay}
  H.V.\ Klapdor-Kleingrothaus \textit{et al.},
  Nucl. Phys. Proc. Suppl. {\bf 100}, 309 (2001);
%
  P.~Benes \textit{et al.}, 
  Phys.\ Rev.\ D {\bf 71}, 077901 (2005)
  [hep-ph/0501295].


\bibitem{tritium}
  V.\ Lobashev \textit{et al.},  
  Nucl. Phys. B (Proc. Suppl.) {\bf 91 }, 280 (2001);
%
  C.\ Weinheimer {\it et al.}, 
  Nucl. Phys. Proc. Suppl. {\bf 118}, 279 (2003). 


\bibitem{nemo2}
  D.~Dassie {\it et al.},
  Nucl.\ Phys.\ A {\bf 678}, 341 (2000).

\bibitem{piondecays}
  D.~I.~Britton {\it et al.},
  Phys.\ Rev.\ D {\bf 49}, 28 (1994);
%
  V.~D.~Barger, W.~Y.~Keung and S.~Pakvasa,
  Phys.\ Rev.\ D {\bf 25}, 907 (1982).


\bibitem{GSR82}
  G.~B.~Gelmini, S.~Nussinov and M.~Roncadelli,
  Nucl.\ Phys.\ B {\bf 209}, 157 (1982).


\bibitem{SNsterile}
  K.~Kainulainen, J.~Maalampi and J.~T.~Peltoniemi,
  Nucl.\ Phys.\ B {\bf 358} (1991) 435.


\bibitem{SNmajoron}
  K.~Choi and A.~Santamaria,
  Phys.\ Rev.\ D {\bf 42}, 293 (1990);
%
  M.~Kachelriess, R.~Tomas and J.~W.~F.~Valle,
  Phys.\ Rev.\ D {\bf 62}, 023004 (2000)
  [hep-ph/0001039];
%
  R.~Tomas, H.~Paes and J.~W.~F.~Valle,
  Phys.\ Rev.\ D {\bf 64}, 095005 (2001)
  [hep-ph/0103017].


\bibitem{delep}
  E.~W.~Kolb, D.~L.~Tubbs and D.~A.~Dicus,
  Astrophys.\ J.\  {\bf 255}, L57 (1982);
%
  G.~M.~Fuller, R.~Mayle and J.~R.~Wilson,
  Astrophys.\ J.\ {\bf 332},  826 (1988);
%
  Z.~G.~Berezhiani and A.~Y.~Smirnov,
  Phys.\ Lett.\ B {\bf 220}, 279 (1989).


\bibitem{sterileBBN}
  K.~Enqvist, K.~Kainulainen and M.~J.~Thomson,
  Nucl.\ Phys.\ B {\bf 373}, 498 (1992);
%
  X.~Shi, D.~N.~Schramm and B.~D.~Fields,
  Phys.\ Rev.\ D {\bf 48}, 2563 (1993) [astro-ph/9307027];
%
  N.~Okada and O.~Yasuda,
  Int.\ J.\ Mod.\ Phys.\ A {\bf 12}, 3669 (1997) [hep-ph/9606411];
%
  S.~M.~Bilenky {\it et al.},
  Astropart.\ Phys.\ {\bf 11}, 413 (1999) [hep-ph/9804421];
%
  P.~Di Bari,
  Phys.\ Rev.\ D {\bf 65}, 043509 (2002);
  Addendum-{\it ibid.}\ D {\bf 67}, 127301 (2003)
  [hep-ph/0108182].


\bibitem{DeltaN}
  S.~Hannestad,
  JCAP {\bf 0305}, 004 (2003) [astro-ph/0303076];
%
  V.~Barger {\it et al.}, 
  Phys.\ Lett.\ B {\bf 566}, 8 (2003) [hep-ph/0305075].


\bibitem{Cyburt:2004yc} 
  R.~H.~Cyburt {\it et al.}, 
  Astropart.\ Phys.\  {\bf 23}, 313 (2005) [astro-ph/0408033].


\bibitem{barbieristerile}
  R.~Barbieri and A.~Dolgov,
  Phys.\ Lett.\ B {\bf 237}, 440 (1990) and
%
  Nucl.\ Phys.\ B {\bf 349}, 743 (1991);
%
  K.~Enqvist, K.~Kainulainen and J.~Maalampi,
  Phys.\ Lett.\ B {\bf 244}, 186 (1990) and
%
  {\it ibid.} B {\bf 249}, 531 (1990);
%
  S.~Dodelson and L.~M.~Widrow,
  Phys.\ Rev.\ Lett.\  {\bf 72}, 17 (1994) [hep-ph/9303287].


\bibitem{Dolgov:1998st}
  A.~D.~Dolgov {\it et al.}, 
  Nucl.\ Phys.\ B {\bf 548}, 385 (1999) [hep-ph/9809598].


\bibitem{FV95}
  R.~Foot and R.~R.~Volkas,
  Phys.\ Rev.\ Lett.\  {\bf 75}, 4350 (1995) [hep-ph/9508275].


\bibitem{BB01}
  K.~S.~Babu and I.~Z.~Rothstein,
 Phys.\ Lett.\ B {\bf 275}, 112 (1992).
%
  L.~Bento and Z.~Berezhiani,
  Phys.\ Rev.\ D {\bf 64}, 115015 (2001) [hep-ph/0108064].


\bibitem{CMBsignal}
  Z.~Chacko {\it et al.}, 
  Phys.\ Rev.\ D {\bf 70}, 085008 (2004) [hep-ph/0312267];
%
  Z.~Chacko {\it et al.}, 
  Phys.\ Rev.\ Lett.\  {\bf 94}, 111801 (2005) [hep-ph/0405067].


\bibitem{BBD04}
  J.~F.~Beacom, N.~F.~Bell and S.~Dodelson,
  Phys.\ Rev.\ Lett.\  {\bf 93}, 121302 (2004) [astro-ph/0404585].


\bibitem{Hannestad04}
  S.~Hannestad,
  JCAP {\bf 0502}, 011 (2005) [astro-ph/0411475].

\end{thebibliography}
\end{document}